\documentclass[twocolumn,aps,prl,superscriptaddress]{revtex4-2}

\usepackage{amsmath}
\usepackage{graphicx}
\usepackage{MnSymbol}
\usepackage{parskip}
\usepackage{makecell}
\usepackage{booktabs}
\usepackage{multirow}
\usepackage{xcolor}

\begin{document}

\title{On the flash temperature in sliding contacts}

\author{M.H. M\"user}
\affiliation{
Dept. of Materials Science and Engineering,
Saarland University, 66123 Saarbr\"ucken, Germany
}

\author{B.N.J. Persson}
\affiliation{Peter Gr\"unberg Institute (PGI-1), Forschungszentrum J\"ulich, 52425, J\"ulich, Germany}
\affiliation{State Key Laboratory of Solid Lubrication, Lanzhou Institute of Chemical Physics, Chinese Academy of Sciences, 730000 Lanzhou, China}
\affiliation{MultiscaleConsulting, Wolfshovener str. 2, 52428 J\"ulich, Germany}

\begin{abstract}
The temperature increase in the contact regions between solids in sliding contact
can easily reach several hundred Kelvin and thereby dramatically affect friction and wear. 
The classical theories by Jaeger, Archard, and Greenwood, commonly used to estimate flash temperature,
ignore the multiscale nature of real surfaces and
 instead approximate the  frictional heat sources with circular or square shapes. 
Here, we present an analytical
theory for the flash temperature valid for randomly rough
surfaces with roughness across arbitrarily many decades in length scale. 
The theory extends established methods for stress correlation functions and peak stresses to 
temperature.
Numerical results for
rubber sliding on concrete, and granite on granite, are presented as illustrations.
We show that classical theories for flash temperature fail severely for surfaces with multiscale roughness.
\end{abstract}

\maketitle

\setcounter{page}{1}
\pagenumbering{arabic}

%\pagestyle{empty}

%%%%%%%%%%%%%% main text %%%%%%%%%%%%%%%%
%\begin{multicols}{2}

%%%%%%%%%%%%%% main text %%%%%%%%%%%%%%%%

{\bf 1 Introduction}

Friction between surfaces generates heat, leading to temperature increases at the contact points. 
This phenomenon is known as flash temperature, which is the high, localized, and brief temperature spike that 
occurs at the true points of contact between two rubbing solids. The rapid generation of heat at these locations 
causes thermal spikes, resulting in intense flash temperatures as kinetic energy is converted into 
heat. These spikes can be extremely high, sometimes reaching 
over $1000\,^\circ {\rm C}$, but they are also incredibly brief,
lasting only for the instant that the asperities are in contact. 
This process is so rapid that the generated heat has little time to conduct away into the bulk of the materials, 
trapping thermal energy and further elevating the temperature at the contact points.

In almost all cases, most of the dissipated energy in sliding friction ends up as thermal energy within the sliding 
solids. The temperature field in the solids can be written as $T({\bf x},t) = T_0({\bf x},t) + \Delta T({\bf x},t)$.
The {\it background} temperature $T_0({\bf x},t)$ varies slowly in space and time while the {\it flash} temperature
$\Delta T({\bf x},t)$, varies rapidly in space and time. 
$\Delta T({\bf x},t)$ is non-zero only close to the asperity contact regions and is highly localized in space.

The flash temperature is nearly independent of the external
conditions, but this is not the case for the background temperature.
If there were no heat transfer to the surrounding environment,
no (time-independent) steady-state temperature profile 
could form, even after a long time. We demonstrate that this holds true even for a semi-infinite system.

Consider a rectangular block sliding on a substrate. We assume that all of
the friction work is converted into heat. If $\sigma ({\bf x},t)$ is the normal stress acting on the block,
we assume that the heat source is $\dot q ({\bf x},t)= \mu v \sigma ({\bf x},t)$, where $v$ is the sliding speed. 
The average $\dot q_0 (t) = \langle \dot q ({\bf x},t)\rangle $ 
(where $\langle .. \rangle$ stands for averaging over the surface area)
determines the background temperature. Assuming no heat transfer to the surroundings
(isolated system) for a semi-infinite system ($z>0$) if sliding starts at
the time $t=0$ the background temperature on the surface $z=0$ at time $t$ is easily obtained from the heat diffusion equation
$$T_0(t) = {1\over \kappa} \left ({D\over \pi}\right )^{1/2} \int_0^t dt' \, {\dot q_0(t') \over (t-t')^{1/2} },\eqno(1)$$
where $\kappa$ is the thermal conductivity and $D$ is the heat diffusivity.
If $\dot q_0$ is time-independent for $t>0$ this gives
$$T_0(t) = {2 \dot q_0 \over \kappa} \left ({D\over \pi}\right )^{1/2}  \sqrt{t}, \eqno(2)$$
which increases without limit as $t\rightarrow \infty$. 
Thus, a steady state (constant temperature distribution) can prevail only if
heat energy is transferred to the surroundings. However, the temperature {\it increase} in the asperity contact regions,
which we will refer to as the flash temperature, reaches a steady state after a very short sliding distance on the of order of the
diameter of the contact regions. For example, a stationary parabolic (Hertzian) heat source 
$$\dot q({\bf x})=\dot q_1 \left [1-\left ({r\over R}\right )^2\right ]^{1/2}, \eqno(3)$$ 
which is turned on at time
$t=0$ results in a time-independent temperature profile for $t \gg R^2/D$, where for $z=0$ and $r<R$ the temperature is
$$\Delta T({\bf x})=T_1 \left [1-{1\over 2} \left ({r\over R}\right )^2\right ] \eqno(4)$$
with $T_1 = \dot q_1 \pi R /4 \kappa$.

\begin{figure*}
\includegraphics[width=0.7\textwidth,angle=0.0]{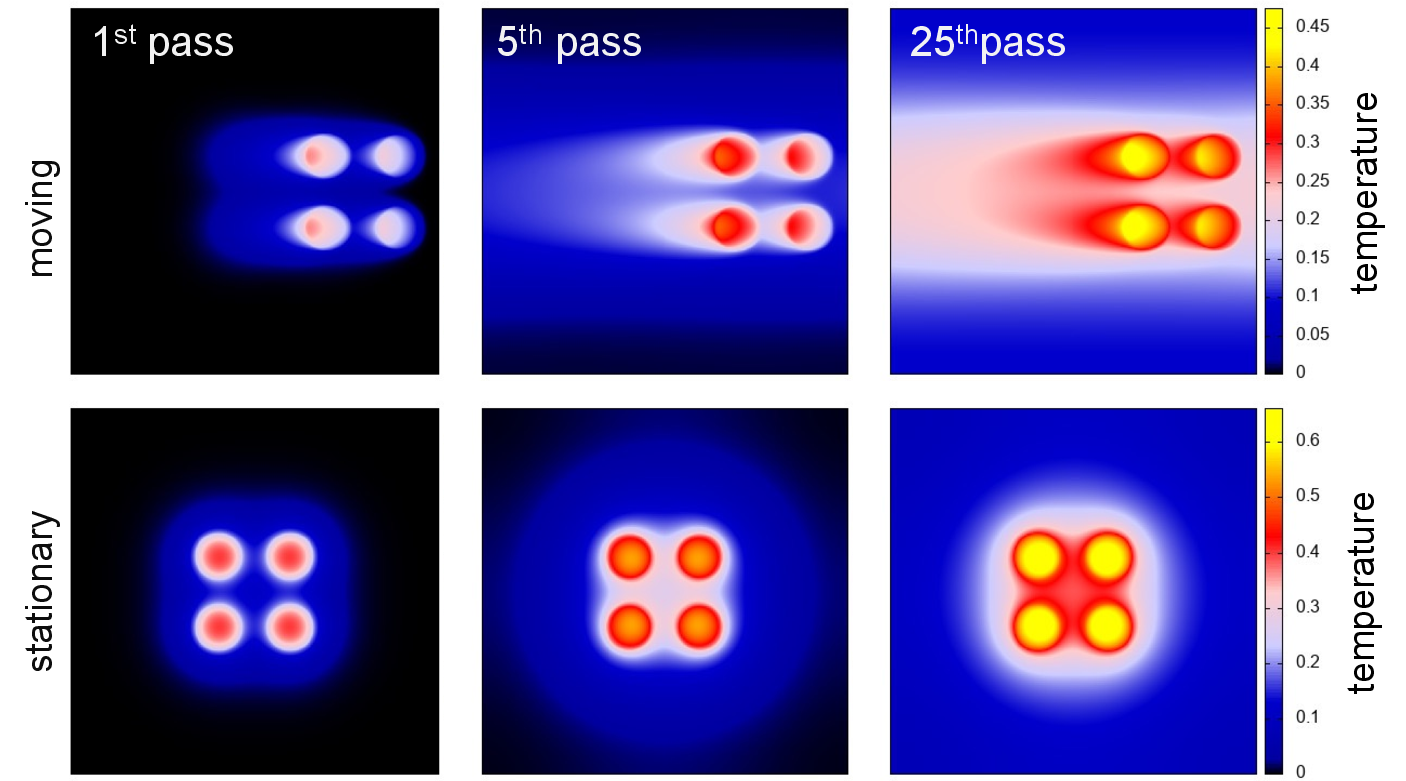}
\caption{\label{MartinFlash.eps}
The temperature on the surface of a semi-infinite solid with a periodic set of heat sources, 
where each unit cell contains four circular heat sources of constant intensity. 
Different panels show the heat distribution at various sliding stages. 
Top row is for $vR/D=20$ and bottom row for $vR/D=0$, where $D$ is the thermal diffusivity.
% The relative temperature difference between the leading edge of the front heat source and 
% its trailing edge is about 0.45 in all cases, with an additional 0.1 increase towards the 
% trailing edge of the back heat source. 
% The final mean temperature offset in the steady state is proportional to the height of the heat-absorbing solid.
}
\end{figure*}

Frictional heating is important in a wide range of applications, spanning from ice and rubber friction to the 
sliding of minerals like granite, which is central to earthquake dynamics.  The flash temperature
occurs in the asperity contact regions, where frictional energy is converted into heat. This localized heating
can have a crucial influence on the resulting friction, usually reducing it. This is the case for sliding ice,
where the flash temperature can melt the ice \cite{ice1,ice2,ice3}, or at least shift the temperature in the contact region towards the
melting point \cite{ice4,ice5}, thereby reducing the frictional shear stress and the friction force.
The same is true for granite 
sliding on granite, where the melting point of the mineral (primarily quartz) may be reached at the sliding speeds (on the order of $\sim 1 \ {\rm m/s}$)
involved in earthquakes \cite{earth1,earth2}. Rubber friction depends exponentially on the temperature. An increase in thermal energy 
shifts the friction coefficient master curve to higher sliding speeds, which usually reduces the friction but sometimes
increases it \cite{rub1,rub2,rub3}.

In Ref. \cite{earth2}, Rice studied the importance of the flash heating to earthquake dynamics.
He considered a model in which the frictional shear stress is constant until the flash temperature 
reaches the order of the pseudotachylyte (rather than quartz or silica) melting point,
after which the shear stress was assumed to be negligible. However, the high stresses and temperatures
in the mineral contact regions already weaken the interface during slip substantially,
well before true melting occurs~\cite{rpp}. Minerals like quartz, which are crystalline, may become amorphous during slip 
in the contact region (where quartz may be locally converted to silica), and will soften continuously 
with increasing temperature and sliding speed \cite{rpp}.
This results in a frictional shear stress that could decrease significantly even before true melting occurs,
as also observed for other crystalline materials, e.g., ice \cite{ice2}, which undergoes cold, displacement-driven 
amorphization \cite{ice4}.

One way the flash temperature can manifest itself in experiments is through
sliding-induced phase transformations in materials.
This was  already observed in the classical studies by Bowden and Tabor, and more recently in many other studies~\cite{mechano0,mechano1}.
However, in general, these transformations may be mechanochemical in nature, involving {\it both} the high contact stresses and the flash temperature.
That is, the chemical or structural modifications involve stress-aided thermal excitations.

All surfaces possess roughness on many length scales, which complicates the determination of the flash temperature.
One key complication is the thermal interaction between the hotspots, which is particularly important between
the closely spaced contact regions within the macroasperity contact areas. 
Classical theories generally treat these heat sources as isolated, neglecting the collective heating effects 
inherent in the hierarchical ``asperity-on-asperity'' structure of real surfaces. 
Alternatively, if they lump many small asperities into larger ones, they do not resolve the local temperature spikes.

The thermal coupling between different scales is illustrated in Fig.~\ref{MartinFlash.eps},
which shows the exact numerical solution of the heat diffusion equation, for four circular and uniform
heat sources on a semi-infinite solid. The top row shows moving heat sources and the lower row shows lower stationary heat sources.
The calculations use periodic boundary conditions so that 
the system consists of a semi-infinite solid with a finite concentration of heat sources. For moving
heat sources at short sliding distances, as in Fig. \ref{MartinFlash.eps}(a), there is negligible thermal overlap between the periodically 
extended heat sources. However, after a long enough
sliding distance, as in Fig.~\ref{MartinFlash.eps}(c), a steady state forms
where the temperature at a heat source depends on the temperature increase induced
by the heat sources in front of it.

In this article, we present a multiscale theory for the flash temperature that includes all relevant length scales.
We derive temperature-temperature and temperature-stress correlation functions, which contain information about
the flash temperature. In the limiting case of roughness on a single length scale, the results are compared to the classical
theories of Jaeger, Archard, and Greenwood \cite{flash1,flash2,flash3} (see also \cite{flash4,flash5,flash6,TianGreen,ReddyhoffHardening,ZhuNumerical}). 
These authors studied the temperature resulting from moving heat sources with circular or rectangular shapes. 
A multiscale contact temperature model was developed by Choudhry, Almqvist, and Larsson \cite{flash6,flashAlm2}, but their framework is purely numerical and is limited 
to surfaces with roughness restricted to a narrow range of wavelengths. The methodology presented here is
analytical and results in equations that can be applied to systems with roughness across arbitrarily many decades of length scales.

In this study, from Sec. 3 and onwards, all temperatures refer to the {\it increase} in the temperature above the background temperature $T_0$.
Thus $T_{\rm flash}$ is a weighted average flash temperature,
and the actual temperature in the contact regions is $T_0+T_{\rm flash}$. Similarly, 
temperature correlation functions like $\langle T({\bf x})T({\bf x}')\rangle$ are calculated with the background temperature
$T_0$ subtracted from $T({\bf x})$. Stated differently, all temperatures refer to actual temperatures only if the background
temperature vanishes.

\vskip 0.2cm
{\bf 2 Preliminary discussion}

Before presenting the formal theory, it is helpful to discuss the different length and time scales entering the model shown 
in Fig.~\ref{MartinFlash.eps} as a conceptual framework for heating in a sliding multi-asperity contact. 
The smallest scale is defined by an individual circular heat source, which, in the given context, is referred to as a micro-asperity with radius $r$.
In the model, the centers of mass of two adjacent micro-asperities are separated by $16 r /5$, yielding an effective radius of $R = (1+8/5) r$ for the macro-asperity. 
The periodically repeated cell is $L = 8r$ long in the in-plane directions and represents a measure of the separation between macro-asperities. 
The height of the cell is $h \approx L$. At the bottom of the cell, the boundary condition $T(x,y,z = -h) = 0$ is imposed.

We first consider the stationary case or, alternatively, a scenario where all heat is dissipated into the sliding body. 
In this case, four distinct relaxation times can be defined, arising from the structure of the homogeneous diffusion equation, 
$${\partial  \over \partial t} T ({\bf x},t) = D \nabla^2 T ({\bf x},t), $$ 
where $D$ is the thermal diffusivity. 
The spatial Fourier transform of this equation, 
$${\partial  \over \partial t} T ({\bf q},t) = -D q^2 T ({\bf q},t), $$ 
connects the relaxation time $\tau_q$ to a temperature undulation with wave vector $q$ via $\tau_q = 1 /D q^2$.

For the micro-asperity, we associate the relaxation with a wave vector $q_r = \pi/r$, yielding $\tau_r = (1/\pi)^2 \approx 0.10$ in units of $[\tau] = r^2/D$. 
Similarly, for the macro-asperity, $q_R = \pi/R$ leads to $\tau_R = (R / \pi r)^2 \approx 0.69$. 
The slowest in-plane relaxation is related to the smallest in-plane wave vector of the periodic cell, $q_L = 2\pi/L$, yielding $\tau_L = (L/2\pi r)^2 \approx 1.62$. 
Finally, for the cell height $h$, the Dirichlet boundary condition at $z=h$ requires a fundamental mode $q_h = \pi/2h$. This results in $\tau_h = (2h/\pi r)^2 \approx 25.9$.
As detailed in the technical comment below, this requires approximately one million time steps to be done for each $\tau_h$ when choosing a mesh size of $a = r/32$.

\begin{figure}[htbp]
\includegraphics[width=0.47\textwidth,angle=0.0]{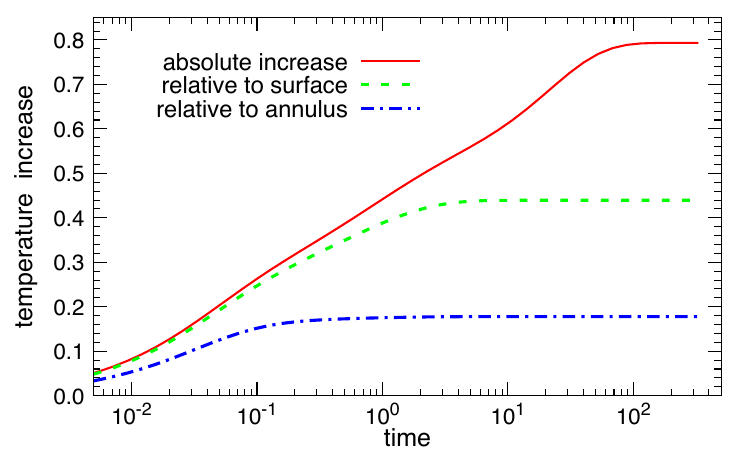}
\caption{\label{relax_static.eps}
Temperature increase at the center point of a microasperity: absolute increase (full, red line), 
relative to the mean surface temperature (dashed, green line), and relative to the asperity's annulus (dash-dotted, blue line).
}
\end{figure}

The dash-dotted, blue line in Fig.~\ref{relax_static.eps} reveals that the local flash temperature, which could be defined as the temperature at the center of a microasperity relative to its annulus, plateaus shortly after the local relaxation time $\tau_r$ has elapsed.
The flash temperature reaches a plateau at a later time when defined relative to the annulus of the macroasperity, or relative to the entire surface.
Since the associated relaxation times, $\tau_R$ and $\tau_L$, are similar in the studied model, we focus on the latter (see the dashed, green line in Fig.~\ref{relax_static.eps}).
All results in later sections will refer to this definition of the flash temperature.
Finally, as is important in practice when the relative contact area is not very small, the absolute flash temperature keeps rising until $\tau_h$ is surpassed.

\begin{figure}
\includegraphics[width=0.47\textwidth,angle=0.0]{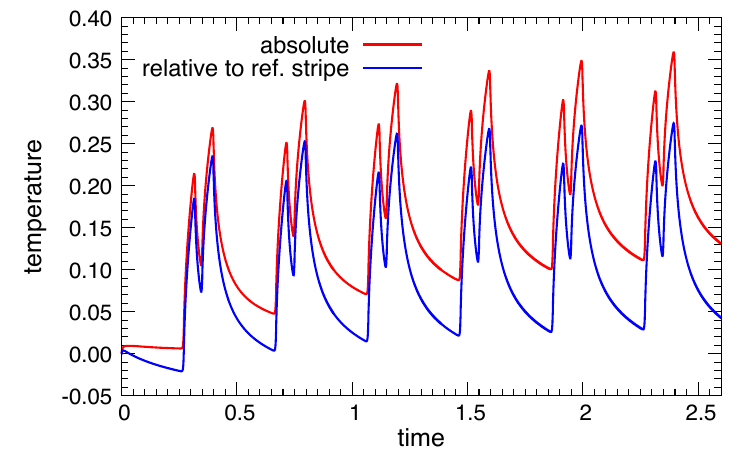}
\caption{\label{relax_time.eps}
Temperature increase at a surface point as a function of time:
absolute increase (solid red line) and increase relative to a reference stripe containing the centers of two circular heat sources (dash-dotted blue line).
}
\end{figure}

When heat is dissipated into the substrate rather than into the slider as illustrated by the thermal footprint shown in Fig.~\ref{MartinFlash.eps} new characteristic time scales emerge. 
Specifically, heat transport on length scales $l$ satisfying $v/l > \tau_l^{-1}$ becomes convective rather than diffusive. 
This transition is evident in Fig.~\ref{relax_time.eps}, where the excess temperature is measured relative to a reference stripe aligned with the centers of two consecutive heat sources. 
The temperature profile assumes nearly the same shape during the first passage of the macro-asperity as it does during the second. 
Thus, the \textit{relative} temperature profile along the sliding direction reaches a quasi-steady state much faster than the absolute temperature.
Each quasi-period of the profile shown here effectively represents the (inverted) spatial temperature distribution along a path through the centers of the micro-asperities, translated into the time domain.

As a technical comment, we remark that the discretization at the interface was $a = L/n_x$, which for $n_x = 512$ corresponds to $a = r/32$.
This yields a maximum wavenumber $q = \pi/a$ with an associated grid-level relaxation time of $\tau_a = (a/\pi)^2 \approx 2.5 \times 10^{-5}$ (in units of $r^2/D$).
For the numerical integration, we employed a forward Euler scheme. 
The integration time step was set to $\Delta t = 0.1 \times (a^2 / D)$, which corresponds to approximately $0.1 \pi^2 \tau_a$. 
This choice is well-suited for the explicit scheme, as it sits safely below the von Neumann stability limit ($\Delta t \le a^2/6D$ for 3D) while providing sufficient temporal resolution to capture the  transients of the micro-asperities.

Yet, $\tau_h/\tau_a \approx 10^6$ time steps must be performed for the largest relaxation time.
While this is still manageable on a $512\times 512$ large system using commodity computers, it illustrates how challenging (non-steady-state) simulations become in a true multiscale system.
Specifically, in our house-written hierarchical solver, where the mesh size is doubled every four layers into the bulk, doubling the $L/a$ ratio increases the required numerical effort 16-fold.
This is because the number of grid points in the outermost layer quadruples while the time step must be reduced by a factor of four. 
In contrast, steady-state profiles can be obtained much more readily, as they require merely one Fast Fourier Transform (FFT) of the heat-flow profile into the Fourier representation, and an inverse FFT to harvest the temperature profile.

\vskip 0.2cm
{\bf 3 Theory}

The temperature distribution in a sliding contact is considered in the steady state after run-in.
We assume that frictional energy is produced only in the 
real contact area within a nanometer-thin surface layer, which we consider to be infinitesimally thin.
We analyze the heat diffusion in the half-space $z>0$ with a heat source (energy per unit area and unit time)
$\dot q ({\bf x},t) = \dot q (x,y,t)$ on the surface $z=0$, and with the heat current $J_z$ vanishing
on the regions of the surface where $\dot q({\bf x},t)=0$. This
heat diffusion problem can be solved most easily by extending the problem to the heat diffusion
in an infinite solid with a heat source on the surface $z=0$, which is twice as strong, $2 \dot q ({\bf x},t)$.
Thus, the temperature distribution in the sliding solid is determined by the heat diffusion equation
$$\rho c_{\rm P} {\partial T \over \partial t} -  \kappa \nabla^2 T  = 2 \dot q ({\bf x},t) \delta (z),$$
where $\rho$, $c_{\rm P}$, and $\kappa$ are the mass density, the specific heat capacity, and the thermal conductivity,
respectively. Introducing the thermal diffusivity $D=\kappa/\rho c_{\rm P}$, we can write
$${\partial T \over \partial t} -  D \nabla^2 T  = {2 D \over \kappa } \dot q ({\bf x},t) \delta (z), \eqno(5)$$
where $\dot q $ depends on the surface position ${\bf x}=(x,y)$ and time $t$. 
The heat source $\dot q$ is non-vanishing only in the area of real contact, which is the area where the normal stress is non-zero.
In the simplest case, one can assume that $\dot q$ is proportional to the normal stress $\sigma ({\bf x})$ so that
$\dot q ({\bf x},t) = \dot q({\bf x}-{\bf v} t)= \mu v \sigma ({\bf x} - {\bf v}t)$.
Writing
$$\dot q ({\bf x},t)  = \int d^2 q \, \dot q({\bf q}) e^{i {\bf q} \cdot ({\bf x}-{\bf v}t)}$$
and using the identity
$$\delta (z) = {1\over 2 \pi} \int dk \, e^{ikz} $$
we obtain from (5)
$$T = {1\over  \pi \kappa } \int d^2q dk  \, {  \dot q ({\bf q}) \over q^2+k^2 
- i{\bf q}\cdot {\bf v}/D} e^{i[{\bf q} \cdot ({\bf x}-{\bf v}t)+kz]}. \eqno(6)$$

We are interested in the temperature at the surface, $z=0$. For this case, the integral in (6) can be performed using the identity
% following standard result
$$\int_{-\infty}^\infty dx \, {1\over x^2 +a^2} = {\pi \over a}.$$
It follows that 
$$T({\bf x},t) = {1\over  \kappa} \int d^2q  \, {  \dot q ({\bf q}) \over \left (q^2
- i{\bf q}\cdot {\bf v}/D \right )^{1/2}} e^{i{\bf q} \cdot ({\bf x}-{\bf v}t)}, $$
or, assuming Amontons' law holds locally,
$$T({\bf x},t) = {\mu v \over  \kappa} \int d^2q  \, {  \sigma ({\bf q}) \over \left (q^2
- i{\bf q}\cdot {\bf v}/D \right )^{1/2}} e^{i{\bf q} \cdot ({\bf x}-{\bf v}t)}. \eqno(7)$$
Writing
$$T({\bf x},t) =  \int d^2q d\omega \, T({\bf q},\omega)  e^{i({\bf q} \cdot {\bf x}-\omega t)},$$
we obtain
$$T({\bf q},\omega) = {\mu v \over  \kappa}{  \sigma ({\bf q}) \over \left (q^2
- i{\bf q}\cdot {\bf v}/D \right )^{1/2}} \delta(\omega - {\bf v} \cdot {\bf q}). \eqno(8)$$

In what follows, we will assume that the average temperature has been subtracted from $T$ so that $\langle T \rangle = 0$.
Similarly, we assume that the average normal stress has been subtracted from $\sigma$ so that $\langle \sigma \rangle = 0$.
This implies that the correlation function $\langle T({\bf x},t) T({\bf 0},0 )\rangle$, when averaged over the surface $z=0$, vanishes,
i.e. the correlation function takes both positive and negative values. The same holds for the stress-stress correlation function.
Note that the average temperature is the ${\bf q}={\bf 0}$ component in (7) so subtracting the average temperature just means excluding the
${\bf q}={\bf 0}$ component in the Fourier expansion of the temperature. The same is true for the normal surface stress.

\begin{widetext}
Using that
$$\langle T({\bf x},t)T({\bf 0},0)\rangle = {(2\pi )^3\over A_0 t_0} \int d^2q d\omega \, \langle T({\bf q},\omega)T(-{\bf q},-\omega) \rangle e^{i({\bf q}\cdot {\bf x}-\omega t)},$$
where $A_0$ is the nominal surface area and $t_0$ is the sliding time, and noting that
$$[\delta(\omega - {\bf v} \cdot {\bf q})]^2 = {t_0\over 2 \pi} \delta(\omega - {\bf v} \cdot {\bf q})$$
we find, using (8):
$$\langle T({\bf x},t)T({\bf 0},0)\rangle = {(2\pi )^2\over A_0} \left ({\mu v \over  \kappa }\right )^2
\int d^2q \,{\langle \sigma({\bf q})\sigma (-{\bf q}) \rangle \over
\left [q^4 +({\bf q}\cdot {\bf v}/D)^2\right ]^{1/2}}e^{i{\bf q}\cdot ({\bf x}-{\bf v} t)} \eqno(9)$$
Furthermore, substituting the stress power spectrum \cite{Persson1}
$$\langle \sigma ({\bf q}) \sigma (-{\bf q})\rangle = {A_0\over (4\pi)^2} (E^*)^2  q^2 C(q) W(q), \eqno(10)$$
where $E^* = E/(1-\nu^2)$ is the effective modulus, 
where $C(q)$ is the surface roughness power spectrum, and
$$W(q) = P(q) [\gamma + (1-\gamma)P^2(q)],$$
where
$$P(q)={\rm erf}\left ({\sigma_0 \over 2 \surd G}\right )$$
$$G= {\pi \over 4} ( E^*)^2 \int_{q_0}^q dq' \ {q'}^3 C(q'),$$
we obtain 
$$\langle T({\bf x},t)T({\bf 0},0)\rangle = \left ({\mu v \over  \kappa }\right )^2 {(E^*)^2\over 4} \int d^2q \, {q^2 C(q) W(q) \over
\left [q^4 +({\bf q}\cdot {\bf v}/D)^2\right ]^{1/2}}e^{i{\bf q}\cdot ({\bf x}-{\bf v} t)}$$
and the equal time correlation function
$$\langle T({\bf x},0)T({\bf 0},0)\rangle = \left ({\mu v \over  \kappa }\right )^2 {(E^*)^2\over 4} \int d^2q \, {q^2 C(q) W(q) \over
\left [q^4 +({\bf q}\cdot {\bf v}/D)^2\right ]^{1/2}}e^{i{\bf q}\cdot {\bf x}}. \eqno(11)$$
The integral over ${\bf q} = (q_x,q_y)$ in 
this expression may be evaluated using a two-dimensional Fast Fourier Transform. Note that the correlation function
depends not only on the distance $r=|{\bf x}|$ but also on the angle $\theta$ between the velocity vector ${\bf v}$
and ${\bf x}$. Here, we focus on the angular averaged (in the $xy$-plane) temperature correlation function, which we denote by $g_T(r)$.
Using the identity
$${1\over 2 \pi} \int_0^{2\pi} d\theta \, e^{iqr {\rm cos}\theta} = J_0 (qr)$$
we obtain
$$g_T(r) = \left ({\mu v \over  \kappa }\right )^2 {(E^*)^2\over 4} \int d^2q \, {q^2 C(q) W(q) \over
\left [q^4 +({\bf q}\cdot {\bf v}/D)^2\right ]^{1/2}} J_0(qr)$$
Transforming to polar coordinates, the ${\bf q}$-integration yields
$$g_T(r) =  \left ({\mu v \over  \kappa}\right )^2 {(E^*)^2\over 4} 
\int_{q_0}^{q_1} dq \, q^2 C(q) W(q) J_0(qr) \int_0^{2\pi} d\phi \,  {1 \over \left [q^2 +(v/D)^2 {\rm cos}^2\phi \right ]^{1/2}}$$
or
$$g_T(r) =  \left ({\mu v \over  \kappa}\right )^2 (E^*)^2
\int_{q_0}^{q_1} dq \, q C(q) W(q) J_0(qr) \int_0^{\pi/2} d\phi \,  {1 \over \left [1 +(v/Dq)^2 {\rm cos}^2\phi \right ]^{1/2}}\eqno(12)$$
The $q$-integral involving the Bessel function $J_0(qr)$ may be evaluated using the method 
described in Appendix D in Ref. \cite{Persson2}.
The $\phi$-integral is of the form 
$$f(x) = \int_0^{\pi /2} d\phi \, {1 \over \left [1 +x \, {\rm cos}^2\phi \right ]^{1/2}}$$
and can be related to the complete elliptic integral of the first kind ($0 \le m < 1$)
% the elliptic integral ($0\le m<1$)
$$K(m) = \int_0^{\pi /2} d\phi \, {1 \over \left [1 -m \, {\rm cos}^2\phi \right ]^{1/2}}$$ 
using
$$f(x) = {1\over (1+x)^{1/2}} K\left ({x\over 1+x}\right ).$$
This representation is computationally advantageous, as highly efficient numerical routines are available for the evaluation of elliptic integrals.

We first consider the two limiting cases of very low and very high velocities.
As $v \rightarrow 0$, Eq.~(12) yields
$$g_T(r) \approx  \left ({\mu v \over  \kappa}\right )^2 (E^*)^2
\int_{q_0}^{q_1} dq \, q C(q) W(q) J_0(qr) {\left( \frac{\pi}{2} \right)}$$
indicating that $g_T \sim v^2$ as $v\rightarrow 0$. To study the $v \rightarrow \infty$ limit, we note that for large $v$:
$$\int_0^{\pi/2} d\phi \,  {1 \over \left [1 +(v/Dq)^2 {\rm sin}^2\phi \right ]^{1/2}} \approx \int_0^{\pi/2} d\phi \,  {1 \over \left [1 +(v/Dq)^2 \phi^2 \right ]^{1/2}} $$
$$\approx {Dq \over v} \int_0^{\pi v / 2 q D} dx \, {1  \over \left [1 +x^2 \right ]^{1/2}} \approx {Dq \over v}  {\rm ln} \left ({2 \pi v \over q D}\right )$$ 
Hence, as $v \rightarrow \infty$,
$$g_T(r) \approx  \left ({\mu v \over  \kappa}\right )^2 (E^*)^2
\int_{q_0}^{q_1} dq \, q C(q) W(q) J_0(qr) {Dq \over v}  {\rm ln} \left ({2 \pi v \over q D}\right )$$
\end{widetext}
showing that $g_T \sim v \, {\rm ln} v$ in the high-velocity limit.

\begin{figure}
\includegraphics[width=0.47\textwidth,angle=0.0]{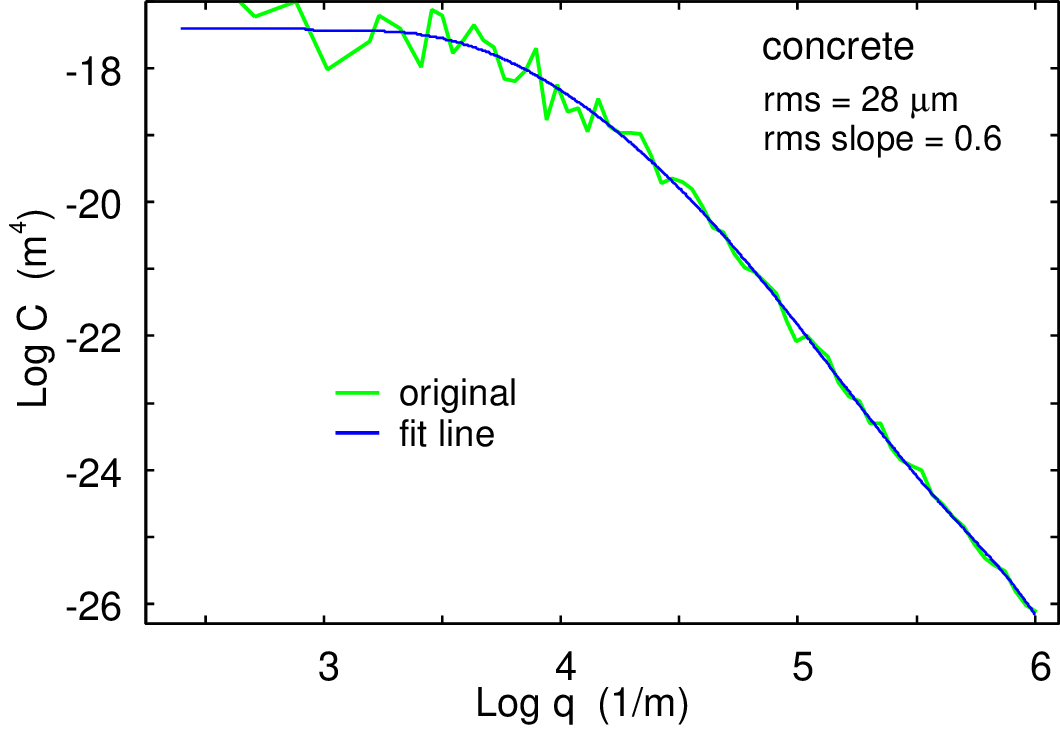}
\caption{\label{1loq.2logC.concrete.eps}
The surface roughness power spectrum of the concrete surface used in the present study. 
The fitted spectrum was employed in the numerical calculations to eliminate non-monotonic 
artifacts in the correlation functions that would otherwise be induced by stochastic scatter in the raw data.
}
\end{figure}

To illustrate the application of the theory, we present numerical results for a representative elastomer system.
We consider a rubber-like solid with 
Young's modulus $E=14 \ {\rm MPa}$ and Poisson ratio $\nu = 0.5$, sliding with a flat bottom surface 
on a rigid substrate.
The substrate roughness is characterized by the power spectrum of a concrete surface, as shown in Fig. \ref{1loq.2logC.concrete.eps}.
The calculations assume a nominal contact pressure $\sigma_0 = 0.1 \ {\rm MPa}$ and 
a friction coefficient $\mu = 1$. The material's thermal properties are defined by a mass density $\rho = 1000$ kg/m$^3$, a specific heat capacity $c_{\rm p} = 1000$ J/(kg·K), and a thermal conductivity $\kappa = 0.1$ W/(m·K).

\begin{figure}
\includegraphics[width=0.47\textwidth,angle=0.0]{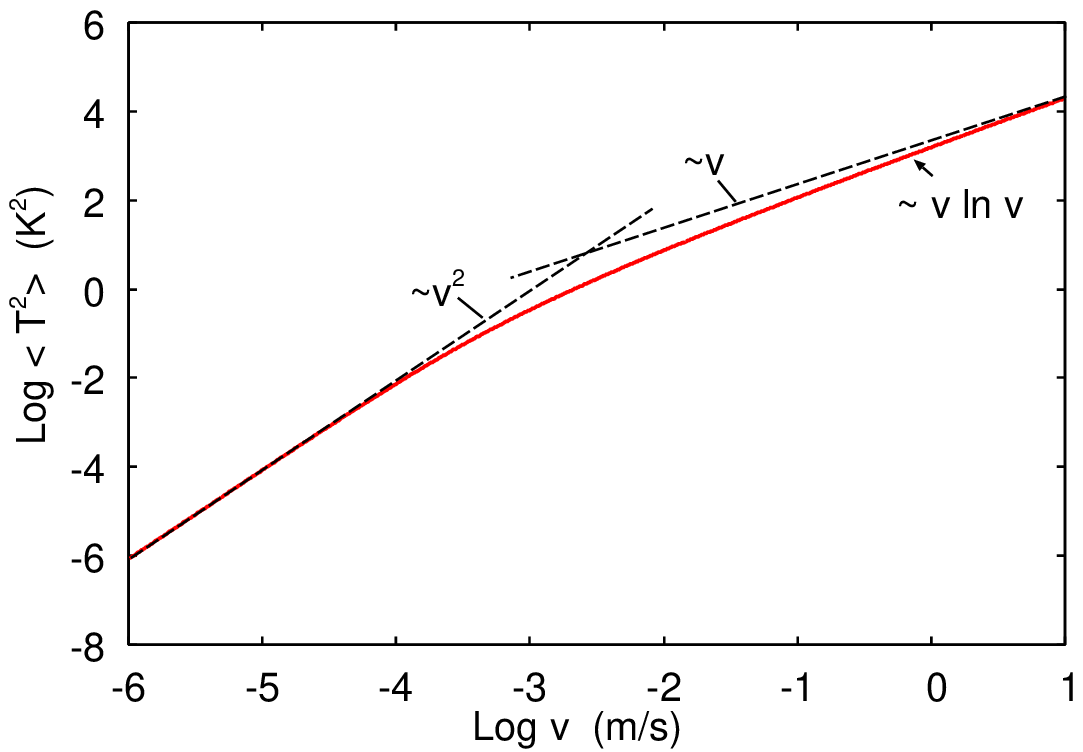}
\caption{\label{1logv.2msTEMPaveraged.eps}
The mean square temperature fluctuation $\langle T^2 \rangle$ as a function of the 
sliding speed (log-log scale). The results are for an elastic solid with 
Young's modulus $E=14 \ {\rm MPa}$ and Poisson ratio $\nu = 0.5$ 
sliding on a rigid concrete surface. The following parameters were assumed: 
nominal contact pressure $\sigma_0 = 0.1 \ {\rm MPa}$, 
friction coefficient $\mu = 1$, mass density $\rho = 1000 \ {\rm kg/m^3}$, 
specific heat capacity $c_{\rm p} = 1000 \ {\rm J/kgK}$, and 
thermal conductivity $\kappa = 0.1 \ {\rm W/Km}$. 
} 
\end{figure}

Fig. \ref{1logv.2msTEMPaveraged.eps}
shows the mean square temperature fluctuation $\langle T^2 \rangle$ as a function of the 
sliding speed on a log-log scale. Note that $\langle T^2 \rangle \sim v^2$ for low sliding speeds
and $\sim v \, {\rm ln} v$ for high sliding speeds. 

\begin{figure}
\includegraphics[width=0.47\textwidth,angle=0.0]{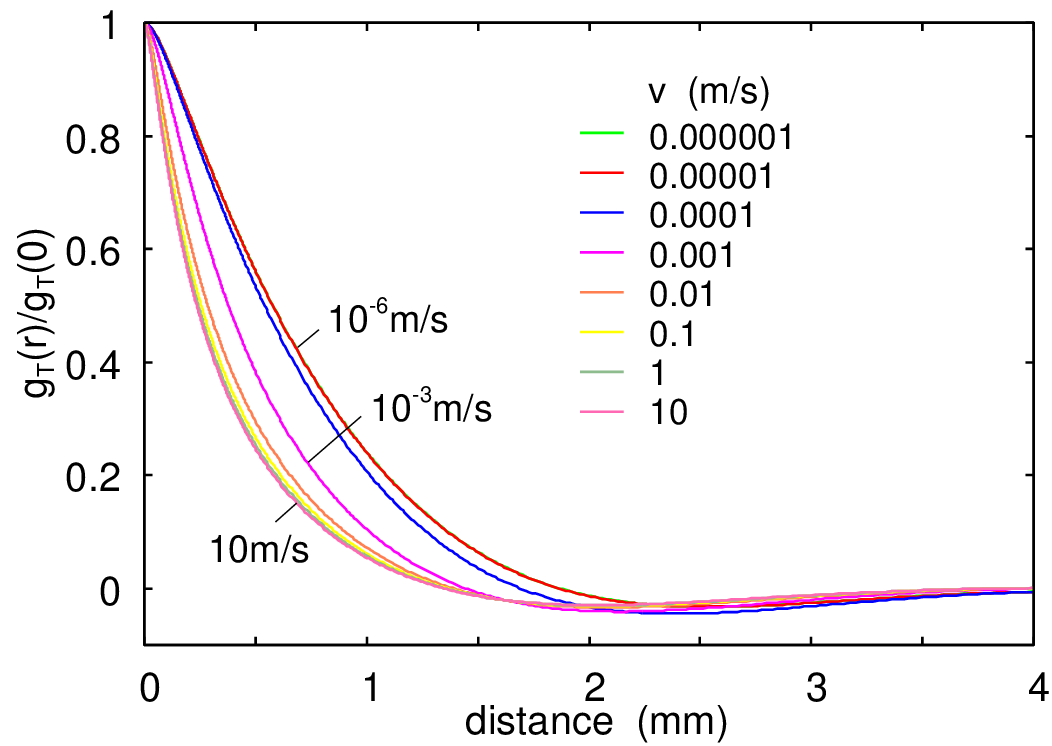}
\caption{\label{1r.2TempCorrelation.vary.velocity.eps}
The dependence of the normalized temperature-temperature correlation function $g_T(r)/g_T(0)$
on the distance $r= |{\bf x}-{\bf x}'|$ for several sliding speeds.
The results are nearly independent of the sliding speed in the high-velocity region ($v > 0.01 \ {\rm m/s}$) 
and the low-velocity region ($v < 10^{-5} \ {\rm m/s}$). These two regions
correspond to the regimes where $\langle T^2 \rangle$ scales with the velocity as
$\sim v \ln v$ and $\sim v^2$, respectively.
}
\end{figure}

Fig. \ref{1r.2TempCorrelation.vary.velocity.eps} 
shows the dependence of the normalized temperature-temperature correlation function $g_T(r)/g_T(0)$
 on the distance $r= |{\bf x}-{\bf x}'|$ for several sliding speeds. 
The results are nearly independent of the sliding speed in the high-velocity region ($v > 0.01 \ {\rm m/s}$) 
and the low-velocity region ($v < 10^{-5} \ {\rm m/s}$). These two velocity regimes 
correspond to the ranges where $\langle T^2 \rangle$ scales with the velocity as 
$\sim v \ln v$ and $\sim v^2$, respectively.

\begin{figure}
\includegraphics[width=0.47\textwidth,angle=0.0]{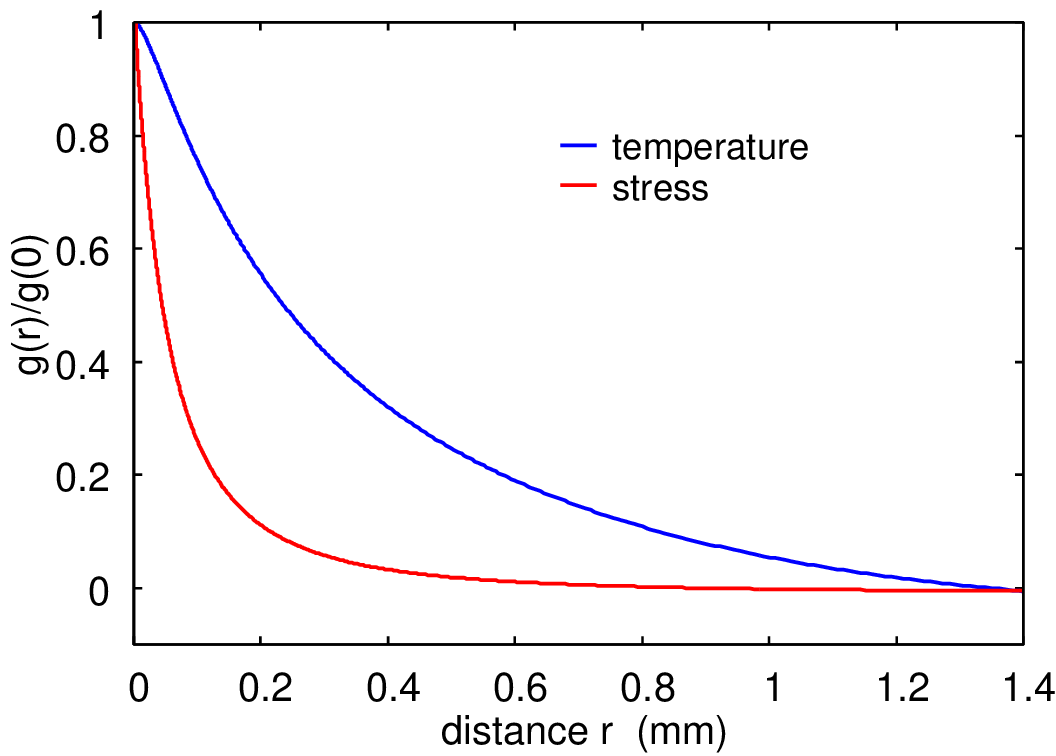}
\caption{\label{1distance.Tcorr.Sig.corr.Tforv10m.p.s.eps}
The dependence of the normalized temperature and stress correlation functions on the 
distance $r= |{\bf x}-{\bf x}'|$. The parameters are the same as in Fig. \ref{1logv.2msTEMPaveraged.eps}. The temperature 
correlation function is calculated for a sliding speed of $v=10 \ {\rm m/s}$. 
}
\end{figure}

\begin{figure}
\centering
\includegraphics[width=0.26\textwidth,angle=0.0]{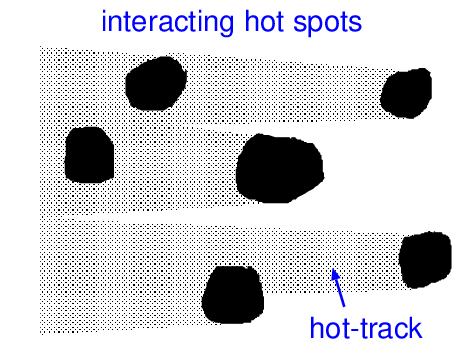}
\caption{\label{HotSpotPic.eps}
At high sliding speeds, a hot track are formed behind each macroasperity contact. 
These tracks overlap with the hotspots and tracks of other asperity contact regions located further downstream.
}
\end{figure}

It is interesting to compare (12) with the stress-stress correlation function
$$g_\sigma (r) = \langle \sigma ({\bf x},0) \sigma({\bf 0},0)\rangle.$$
Using (10), we obtain \cite{Persson2,Persson3}:
$$g_\sigma (r) = (E^*)^2 {\pi \over 2} \int_{q_0}^{q_1} dq \, q^3 C(q) W(q) J_0(qr). \eqno(13)$$
In Fig. \ref{1distance.Tcorr.Sig.corr.Tforv10m.p.s.eps}, 
we compare the dependence of $g_T(r)/g_T(0)$ and $g_{\sigma} (r)/ g_{\sigma}(0)$ on the 
distance $r= |{\bf x}-{\bf x}'|$. The temperature 
correlation function is calculated for a sliding speed of $v=10 \ {\rm m/s}$.
The radius of the nominal contact region can be deduced from the stress-stress correlation function 
and is $r_0 < 1 \ {\rm mm}$. At the sliding speed $v$, it takes a time $\tau = r_0/v$ to move 
a distance on the order of the contact region size. During this time, the heat diffuses 
a distance $d \approx (D\tau)^{1/2} \approx (D r_0/v )^{1/2} \approx 3 \ \mu{\rm m}$, 
which is much smaller than the size of the asperity contact region. Nevertheless, 
Fig. \ref{1distance.Tcorr.Sig.corr.Tforv10m.p.s.eps} shows that the effective width of the 
temperature distribution is much larger than that of the stress distribution. The reason for this is 
the hot tracks left on the rubber surface behind the moving contact regions (see Fig. \ref{HotSpotPic.eps}). 
At high sliding speeds, these hot track will only disappear (via heat diffusion) at a long distance behind the 
moving contact regions.

Since the temperature distribution extends far beyond the contact regions while the stress correlation is non-vanishing only 
within the contact area, the normalized temperature distribution is more affected than the stress distribution by a change 
in the nominal contact pressure. We will illustrate this with two examples.

\begin{figure}
\includegraphics[width=0.47\textwidth,angle=0.0]{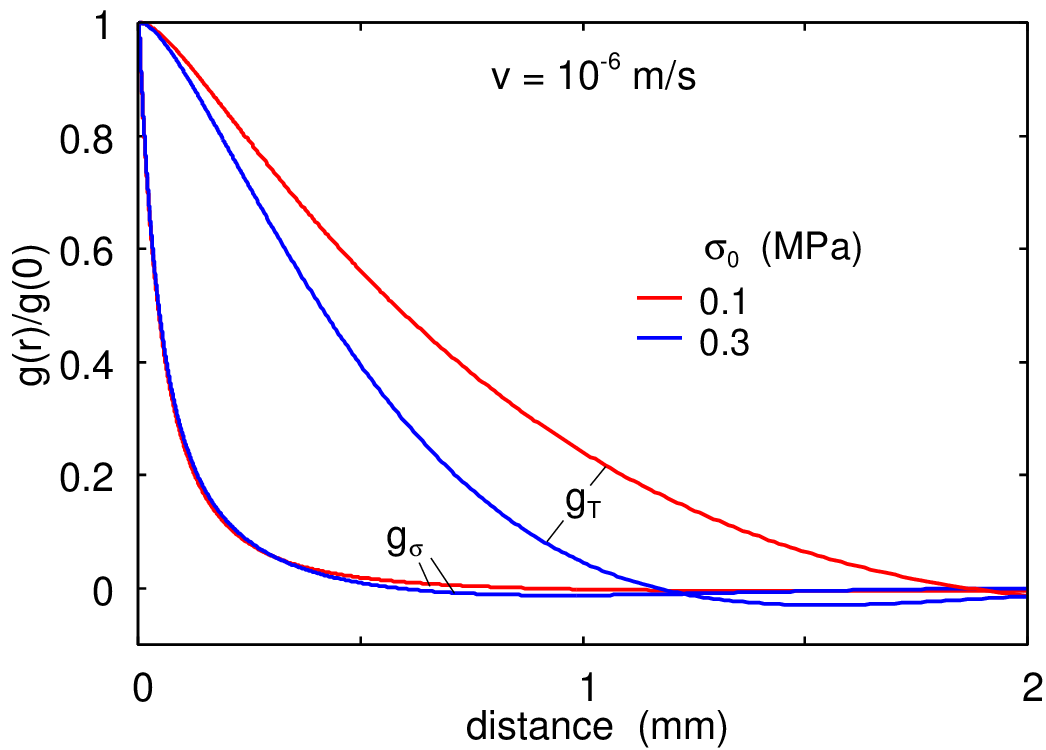}
\caption{\label{1distance.2Tcorr.Sigcorr.Em06.m.p.s.Sig.0.1MPa.0.3MPa.eps}
The dependence of the normalized temperature and stress correlation functions on the 
distance $r= |{\bf x}-{\bf x}'|$ for nominal pressures $\sigma_0 = 0.1$ MPa (red curves) 
and $0.3$ MPa (blue curves). The parameters are the same as in Fig. \ref{1logv.2msTEMPaveraged.eps}. The temperature 
correlation function is calculated for a low sliding speed of $v=10^{-6}$ m/s. 
%\textcolor{blue}{Note that without normalization, but including the temperature offset, 
%the high-pressure (blue) curves would lie significantly above the low-pressure (red) curves for every value of $r$.}
}
\end{figure}

\begin{figure}
\includegraphics[width=0.47\textwidth,angle=0.0]{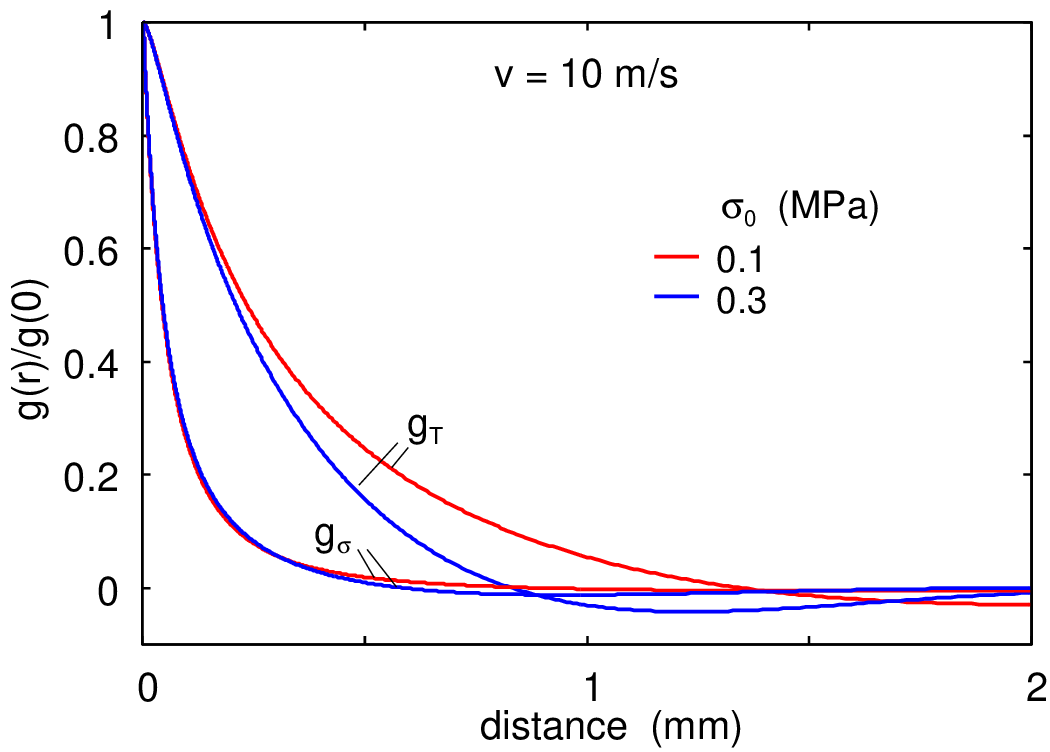}
\caption{\label{1distance.2Tcorr.Sigcorr.10m.p.s.Sig.0.1MPa.0.3MPa.eps}
The dependence of the normalized temperature and stress correlation functions on the 
distance $r= |{\bf x}-{\bf x}'|$ for nominal pressures $\sigma_0 = 0.1$ MPa (red curves) 
and $0.3$ MPa (blue curves). The parameters are the same as in Fig. \ref{1logv.2msTEMPaveraged.eps}. The temperature 
correlation function is calculated for a high sliding speed of $v=10$ m/s. 
}
\end{figure}

As the contact pressure increases, the concentration of macroasperity contact 
regions increases. For low sliding speeds, heat diffusion broadens the temperature profile, and the profile 
approaches the average temperature faster with increasing $r$ as the concentration of macroasperities grows. 
Similarly, for high sliding speeds, the tails of the hot tracks (see Fig. \ref{HotSpotPic.eps}) overlap 
other contact regions more frequently; consequently, the temperature profile again approaches the average temperature 
faster as $\sigma_0$ increases. As a result, in both cases, 
the normalized temperature correlation will decay faster as the contact pressure increases. 
This is illustrated in Fig. \ref{1distance.2Tcorr.Sigcorr.Em06.m.p.s.Sig.0.1MPa.0.3MPa.eps} 
for low sliding speeds and in Fig. \ref{1distance.2Tcorr.Sigcorr.10m.p.s.Sig.0.1MPa.0.3MPa.eps} for high sliding speeds.

In applications to sliding friction, the temperature within the asperity contact regions is of primary importance, 
rather than the temperature distribution across the entire surface. To isolate this, one can define an effective 
flash temperature by weighting the temperature by the local contact stress:
$$T_{\rm flash} = \frac{\langle T({\bf x}) \sigma ({\bf x}) \rangle}{\langle \sigma ({\bf x}) \rangle}, $$
where $\langle \sigma ({\bf x}) \rangle = \sigma_0$ is the nominal contact pressure. 
Defined this way, the temperature is weighted only where the stresses are high, and is ignored in the non-contact regions 
where $\sigma(\mathbf{x}) = 0$.
Using the relation
$$\langle T({\bf x}) \sigma ({\bf x}) \rangle 
= \frac{(2\pi )^2}{A_0} \int d^2q \, \langle T({\bf q}) \sigma (-{\bf q}) \rangle $$
and substituting (8), we obtain:
$$\langle T({\bf x}) \sigma ({\bf x}) \rangle  = \frac{\mu v}{\kappa} \frac{(2\pi )^2}{A_0} \int d^2q \, 
\frac{\langle \sigma ({\bf q}) \sigma (-{\bf q}) \rangle}{(q^2 -i {\bf q}\cdot {\bf v}/D)^{1/2}}.$$
Finally, applying (10) yields the expression:
\begin{widetext}
$$ T_{\rm flash} = \frac{\mu v}{\kappa \sigma_0} (E^*)^2 \int_{q_0}^{q_1} dq \, q^2 C(q) W(q) 
{\rm Re} \int_0^{\pi/2} d\phi \, \frac{1}{[1 -i (v/Dq) \cos\phi ]^{1/2}}. \eqno(14)$$

The temperature distribution within the macroasperity contacts during sliding is anisotropic, with higher 
temperatures occurring at the trailing edge compared to the leading edge of the contact. 
This asymmetry arises because the material at the trailing edge has already been heated by the preceding contact area. 
The anisotropy of the average temperature distribution can be quantified using the approach described above. 
We consider the effective temperature gradient:
$$ \nabla T_{\rm flash} = \frac{\langle \nabla T({\bf x}) \sigma ({\bf x})\rangle}{\langle \sigma ({\bf x}) \rangle}$$
Using the relation
$$\nabla T({\bf x}) = \nabla \int d^2q \, T({\bf q}) e^{i {\bf q} \cdot {\bf x}} = \int d^2q \, (i {\bf q}) T({\bf q}) e^{i {\bf q} \cdot {\bf x}},$$
we obtain:
$$\langle \nabla T ({\bf x}) \sigma ({\bf x}) \rangle 
= \frac{(2\pi )^2}{A_0} \int d^2q \, (i{\bf q}) \langle T({\bf q}) \sigma (-{\bf q}) \rangle = \frac{\mu v}{\kappa} \frac{(2\pi )^2}{A_0} \int d^2q \, \frac{(i {\bf q}) 
\langle \sigma ({\bf q}) \sigma (-{\bf q}) \rangle}{(q^2 -i {\bf q}\cdot {\bf v}/D)^{1/2}}. \eqno(15)$$
By choosing the $x$-axis along the sliding direction, only the $x$-component of (15) remains non-vanishing. 
Denoting this as $T'_{\rm flash}$ (where the prime indicates the derivative with respect to $x$), we find:
$$ T'_{\rm flash} = \frac{\mu v}{\kappa \sigma_0} (E^*)^2 \int_{q_0}^{q_1} dq \, q^3 C(q) W(q) 
{\rm Re} \int_0^{\pi/2} d\phi \, \frac{i \cos\phi}{[1 -i (v/Dq) \cos\phi ]^{1/2}}. \eqno(16)$$

Higher-order derivatives of $T({\bf x})$ can be calculated in a similar manner. 
For instance, considering that a Laplacian in real space is just a multiplication by $-q^2$ in Fourier space, we obtain
$$\langle \nabla^2 T ({\bf x}) \sigma ({\bf x}) \rangle 
= \frac{\mu v}{\kappa} \frac{(2\pi )^2}{A_0} \int d^2q \, \frac{(-q^2) 
\langle \sigma ({\bf q}) \sigma (-{\bf q}) \rangle}{(q^2 -i {\bf q}\cdot {\bf v}/D)^{1/2}}.$$
Defining the effective flash temperature Laplacian as
$$ \nabla^2 T_{\rm flash} = \frac{\langle \nabla^2 T ({\bf x}) \sigma ({\bf x}) \rangle}{\langle \sigma ({\bf x}) \rangle },$$
we find:
$$ \nabla^2 T_{\rm flash} = -\frac{\mu v}{\kappa \sigma_0} (E^*)^2 \int_{q_0}^{q_1} dq \, q^4 C(q) W(q) 
{\rm Re} \int_0^{\pi/2} d\phi \, \frac{1}{[1 -i (v/Dq) \cos\phi ]^{1/2}}. \eqno(17)$$
\end{widetext}

\begin{figure}
\includegraphics[width=0.47\textwidth,angle=0.0]{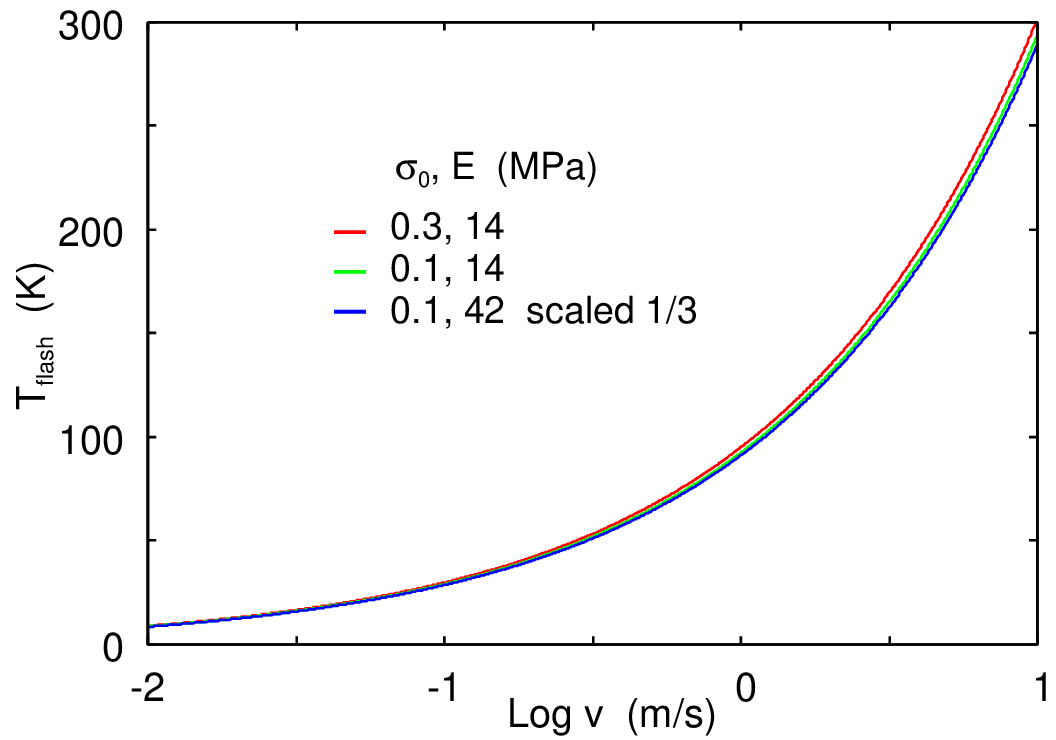}
\caption{\label{1logv.2FlashTemp.red.0.3MPa.14MPa.green.0.1MPa.14MPa.blue.0.1MPa.42MPa.eps}
The flash temperature $T_{\rm flash}$ as a function of the sliding speed for an elastic block sliding on a concrete surface. 
The red and green lines correspond to nominal contact pressures of $\sigma_0 = 0.3$ MPa and $0.1$ MPa, respectively, 
with a Young's modulus of $E=14$ MPa. The blue line is for $\sigma_0 = 0.1$ MPa and $E=42$ MPa, 
but with the flash temperature scaled by $1/3$. In all calculations, the Poisson ratio $\nu = 0.5$, 
friction coefficient $\mu = 1$, mass density $\rho = 1000 \ {\rm kg/m^3}$, 
heat capacity $c_{\rm p} = 1000 \ {\rm J/kgK}$, and thermal conductivity $\kappa = 0.1 \ {\rm W/mK}$.
}
\end{figure}

We now present numerical results for the flash temperature. 
Fig. \ref{1logv.2FlashTemp.red.0.3MPa.14MPa.green.0.1MPa.14MPa.blue.0.1MPa.42MPa.eps} 
shows $T_{\rm flash}$, obtained from (14), 
as a function of sliding speed for an elastic block on the concrete surface. 
The red and green lines represent contact pressures of $\sigma_0 = 0.3$ MPa and $\sigma_0 = 0.1$ MPa, 
respectively, both with a Young's modulus of $E=14$ MPa. 
The blue line corresponds to $\sigma_0 = 0.1$ MPa and $E=42$ MPa, with the flash temperature scaled by a factor of $1/3$.

Note that the result is independent of the nominal contact pressure. 
This holds as long as the pressure is sufficiently small such that the relative contact area $A/A_0 \ll 1$.  In this regime, 
increasing the pressure does not change the size or pressure distribution within individual contact regions, 
but merely increases the number of macroasperity contacts proportional to $\sigma_0$. Furthermore, lateral thermal interaction between hotspots remains negligible if the separation between contact regions is large.

Similarly, when $A/A_0 \ll 1$, increasing the Young's modulus does not significantly change the 
size of the macroasperity contact regions, but 
decreases their number proportional to $1/E$. Consequently, the 
local pressure acting within those contact regions increases proportional to $E$, resulting in a flash 
temperature that is also proportional to $E$. This explains why scaling the $E=42$~MPa curve by $1/3$ 
yields the same temperature-velocity dependence as the $E=14$~MPa case.

\begin{figure}
\includegraphics[width=0.47\textwidth,angle=0.0]{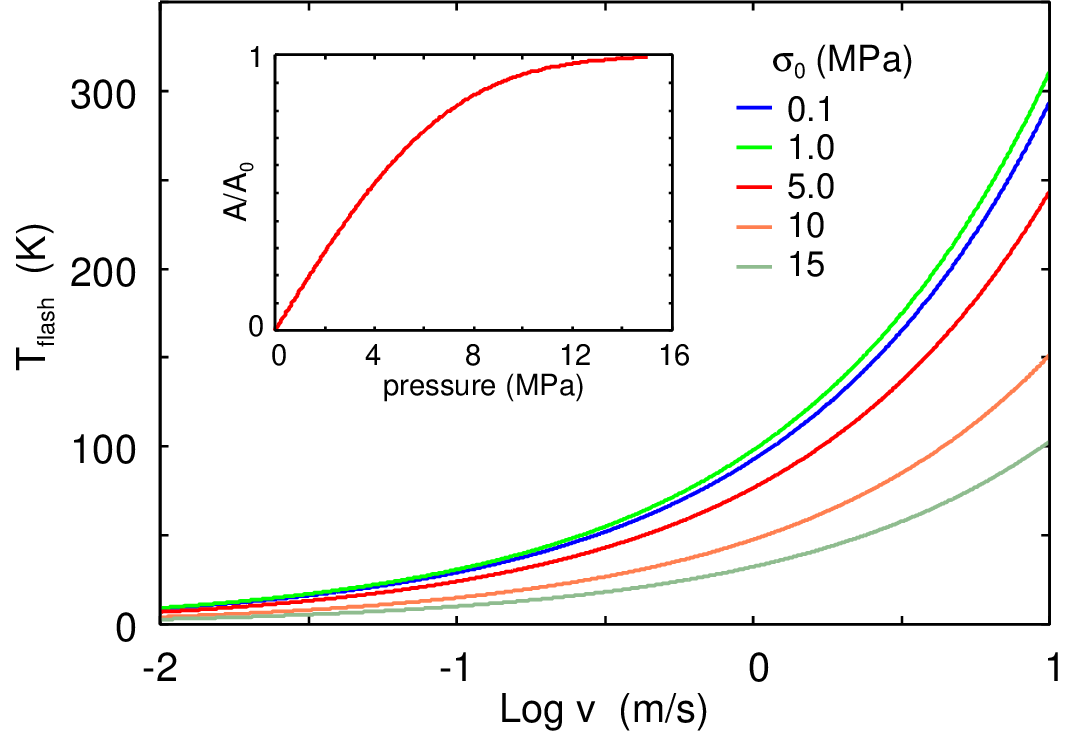}
\caption{\label{1logv.2flash.ManyPressuresRubber.eps}
The flash temperature  for different nominal contact pressures indicated.
The inset shows the contact area as a function of the applied pressure $\sigma_0$.
%Note that the flash temperature decreases as $\sigma_0$ increases. At the same time the background temperature (not shown),
%which is the cumulative contribution from the flash temperature, will increase. The reason the flash temperature decreases
%is easy to understand in the limit where the frictional heat source $\dot q$ is constant in the contact regions. In this case
%the average surface temperature averaged over the whole surface will approach the temperature in the contact area as
%$A/A_0$ approach 1 so in that limit the flash temperature vanish and the surface temperature will be constant on the whole surface.
}
\end{figure}

When the applied pressure $\sigma_0$ increases the flash temperature $T_{\rm flash}$ decreases,
see Fig. \ref{1logv.2flash.ManyPressuresRubber.eps}.
At the same time the background temperature (not shown),
which is the cumulative contribution of the flash temperature, increases in such a way that the surface temperature everywhere
on the surface increases. The reason the flash temperature decreases
is easy to understand in the special case where the frictional heat source $\dot q$ is constant in the contact regions. In this case
the surface temperature averaged over the whole surface will approach the temperature in the contact area as
$A/A_0 \rightarrow 1$ so in that limit the flash temperature vanish and the surface temperature will be constant 
on the whole surface, and given by (2).

\vskip 0.2cm
{\bf 4 One length-scale roughness: comparing the theory with the classical result}

It is interesting to compare the expression (14) with the standard expression for the maximum flash temperature resulting from a circular uniform heat source 
(a disc with radius $r_{\rm c}$) moving with velocity $v$ on the surface of a semi-infinite solid. For this case Greenwood have presented an interpolation formula:
$$\Delta T \approx {\sigma_{\rm f} v \alpha r_{\rm c} \over \kappa} {1 \over [1+ (\alpha /\beta)^2 \pi r_{\rm c} v/ 8  D ]^{1/2}}. \eqno(18)$$
When $\alpha=\beta=1$ this equation interpolates between the known {\it maximum} flash temperature for low and high sliding velocities,
and with $\alpha \approx 0.849$ and $\beta \approx 0.610$ it interpolation between the known {\it average} temperature for low and high velocities.
The expression (14) presents a weighted average of the contact temperature so when compared to (18) we expect to find $\alpha < 1$ and $\beta <1$.

Using the local friction stress, $\sigma_{\rm f} = \mu \bar \sigma$, we obtain
$$\Delta T \approx {\mu \bar \sigma v \alpha r_{\rm c} \over \kappa} {1 \over [1+ (\alpha /\beta)^2 \pi r_{\rm c} v/ 8 D ]^{1/2}} \eqno(19)$$
The classical approach assumes roughness only on a single length scale, where the contact regions are compact.
This corresponds to using a power spectrum of the form:
$$C(q) \approx C_0 \delta (q-q_{\rm m}), \eqno(20)$$
where $C_0$ is chosen such  that the rms slope equals $\xi_{\rm m}$:
$$\xi_{\rm m}^2 = 2\pi \int_0^\infty dq \,q^3 C(q) = 2 \pi q_{\rm m}^3 C_0,$$
which yields
$$C_0 = {\xi_{\rm m}^2  \over 2 \pi q_{\rm m}^3 }\eqno(21)$$
Substituting (20) and (21) into (14) gives
\begin{widetext}
$$ T_{\rm flash} = {\mu v \over \kappa \sigma_0} (E^*)^2  q_{\rm m}^2 {\xi_{\rm m}^2  \over 2 \pi q_{\rm m}^3 } W(q_{\rm m}) 
{\rm Re} \int_0^{\pi/2} d\phi \, {1\over [1 -i (v/Dq_{\rm m}) {\rm cos}\phi ]^{1/2}}.$$

The classical approach assumes well separated asperity contact regions, in which case the thermal interaction between the contact regions can be neglected.
This corresponds to the limit, where the relative contact area $A_{\rm m}/A_0 \ll 1$.
In this regime, $W(q_{\rm m}) \approx \gamma A_{\rm m}/A_0$ with $\gamma \approx 0.5$, so that
$$ T_{\rm flash} \approx {\mu v \over \kappa \sigma_0} (E^*)^2  {\gamma \xi_{\rm m}^2  \over 2 \pi q_{\rm m} } {A_{\rm m}\over A_0}
{\rm Re} \int_0^{\pi/2} d\phi \, {1\over [1 -i (v/Dq_{\rm m}) {\rm cos}\phi ]^{1/2}}. \eqno(22)$$
%\end{widetext}
At small loads, when $ A_{\rm m}/A_0 \ll 1$, we have \cite{Mark,Persson,Martin}
$${A_{\rm m}\over A_0} \approx {2 \sigma_0 \over \xi_{\rm m} E^*}$$
so that 
$$ T_{\rm flash} \approx {\mu v \over \kappa \sigma_0} {\gamma  \over 2 \pi q_{\rm m} } 4 \sigma_0^2 {A_0\over A_{\rm m}}
{\rm Re} \int_0^{\pi/2} d\phi \, {1\over [1 -i (v/Dq_{\rm m}) {\rm cos}\phi ]^{1/2}}.$$
Using that $\bar \sigma = \sigma_0 A_0 /A_{\rm m}$ yields
$$ T_{\rm flash} \approx {\mu  \bar \sigma v \over \kappa} {\gamma  \over q_{\rm m} } 
{2\over \pi} {\rm Re} \int_0^{\pi/2} d\phi \, {1\over [1 -i (v/Dq_{\rm m}) {\rm cos}\phi ]^{1/2}}.\eqno (23)$$
\end{widetext}
The function
$$f(x) = {2\over \pi} {\rm Re} \int_0^{\pi/2} d\phi \, {1\over [1 -i x {\rm cos}\phi ]^{1/2}}$$
has the limits $f \rightarrow 1$ as $x\rightarrow 0$ and $f\rightarrow 2/{\sqrt{\pi x}}$ % \surd (\pi x)$ 
as $x \rightarrow \infty $. Thus as $v \rightarrow 0$, we get from (23):
$$ T_{\rm flash} \approx {\mu  \bar \sigma v \over \kappa} {\gamma  \over q_{\rm m} },$$
while in the same limit the classical expression (19) gives 
$$\Delta T \approx {\mu \bar \sigma v \alpha r_{\rm c} \over \kappa}. $$
Thus, if we choose 
$$r_{\rm c} = {\gamma \over \alpha q_{\rm m}}\eqno(24)$$ 
the results agree. As $v \rightarrow \infty $ Eq. (23) gives
$$T_{\rm flash} \approx {\mu  \bar \sigma v \over \kappa} {\gamma  \over q_{\rm m} } \left ({4 D q_{\rm m} \over \pi v} \right )^{1/2},$$
while (19) gives
$$\Delta T \approx {\mu \bar \sigma v \beta r_{\rm c} \over \kappa} \left ({8D \over \pi r_{\rm c} v}\right )^{1/2}$$
These two expressions agree if 
$$r_{\rm c} = {\gamma^2 \over 2 \beta^2 q_{\rm m}}$$
Using (24) this imply $\beta = (\alpha \gamma /2)^{1/2} \approx 0.5 \alpha^{1/2}$. 

%\begin{widetext}
%
\begin{table*}[hbt]
   \caption{The mass density $\rho$, the heat conductivity $\kappa_{\rm th}$, the heat capacity per unit mass $C_V$, and the heat diffusivity $D$
for different materials.}
   \label{thermal}
   \renewcommand{\arraystretch}{1.5} % Default value: 1
   \begin{center}
      \begin{tabular}{@{}|l||c|c|c|c|@{}}
         \hline
            solid   &  $\rho \ {\rm (kg/m^3)}$  & $\kappa_{\rm th} \ {\rm (W/Km)}$ & $C_V \ {\rm (J/kgK)}$ & $D \  {\rm (m^2/s)}$ \\
         \hline
         \hline
            aluminum & 2700 & 240 & 490 & $1.8\times 10^{-4}$ \\
         \hline
            steel & 7900 & 10-50 & 490 & $(2.6-12.9) \times 10^{-6}$ \\
         \hline
            rubber  & 1200 & 0.2-0.6 & 1600 & $(1.0-2.6)\times 10^{-7}$ \\
         \hline
            silica  & 2400 & 1.4 & 700 & $8.3 \times 10^{-7}$  \\
         \hline
            quartz  & 2400 & 11.7 & 700 & $7.0 \times 10^{-6}$  \\
         \hline
      \end{tabular}
   \end{center}
\end{table*}
%
%\end{widetext}

\vskip 0.2cm
{\bf 5 Qualitative discussion and more numerical results}

Heat diffusion is characterized by the heat diffusivity $D=\kappa /(\rho C_{\rm p})$, where $\kappa$, $\rho$, and $C_{\rm p}$
are the heat conductivity, mass density, and heat capacity, respectively. In Table \ref{thermal}, we give the heat diffusivity at room temperature for some 
materials of interest. Most materials have a heat diffusivity $D$ between $10^{-7}$ and $10^{-4}$ m$^2$/s.
Qualitatively, the heat diffusion equation shows that during the time $\tau$, heat diffusion will broaden the 
temperature distribution by a characteristic distance $d \approx (D \tau)^{1/2}$.

A macroasperity contact region consists of many smaller, closely spaced microasperity
contact regions. The microasperity contact regions consist, in turn, of even smaller 
asperity contact regions. 
We may say that the contact area in a macroasperity contact region exhibits granularity on many length
scales, which could extend down to atomistic scales. The frictional heat source will show the same granularity as the
contact area. The flash temperature in one microasperity contact region depends not only on the frictional energy dissipation in the contact
area but the local flash temperature will be enhanced from the heat diffusion or convection from the other 
microasperity contact areas in the same macroasperity contact region.

The microasperity contact regions within a macroasperity contact area are densely distributed. This is the case even when the nominal
contact pressure is (arbitrary) small, where the separation between the macroasperity contact regions are (arbitrary) large.
Hence, even if the thermal coupling between the macroasperity contact regions can be neglected, this is never the case for the microasperity contact regions
within the macroasperity contact regions. At low sliding speed thermal diffusion will couple the microasperity heat spots. For high sliding
speeds a microasperity contact close to the exit of the macroasperity contact region will experience the hot tracks from the microasperity contact
regions in front of it.

\begin{figure}
\includegraphics[width=0.47\textwidth,angle=0.0]{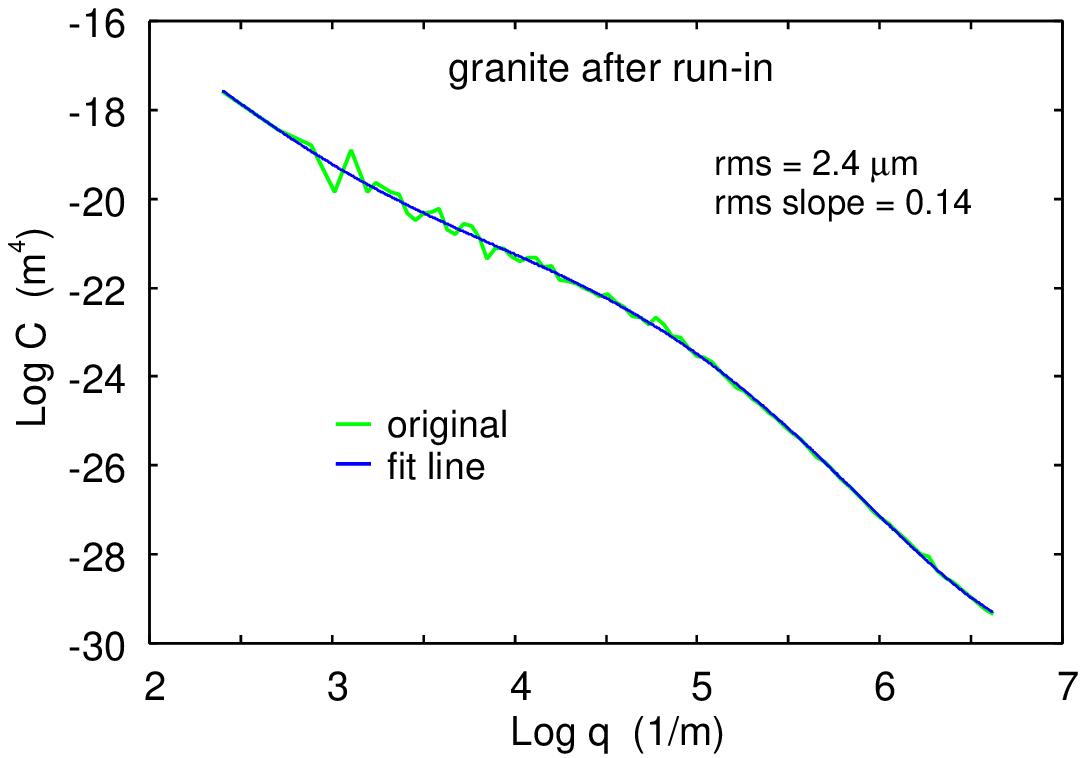}
\caption{\label{1logq.2logC.granite.after.runin.eps}
The surface roughness power spectrum of the granite surface used in the present study. The spectrum represents the run-in state and is shown as a function of the wavenumber $q$.
}
\end{figure}

The numerical results presented in Sec. 2 correspond roughly to those characteristic of rubber sliding on a 
concrete surface. Here, we will extend the study to quartz-on-quartz and silica-on-silica, 
which are minerals of interest in earthquake dynamics, as the continental parts of tectonic 
plates are often described as \textit{granitic} in overall composition, and granite consists mainly of quartz.
We will use the measured surface roughness power spectrum of a run-in granite block and substrate
system shown in Fig. \ref{1logq.2logC.granite.after.runin.eps} (see Ref. \cite{rpp}). 
After run-in, the contact regions deform mostly elastically at the length scales for which
the power spectrum has been measured (down to the micrometer length scale, which is the resolution of the stylus topography-instrument used). Thus, we will neglect plastic flow in the model calculations and will use the thermal parameters
for quartz and silica given in Table~\ref{thermal}. 

Assume that a high asperity from one of the granite surfaces, say the sliding block, contacts the other granite surface.
In that case, it is likely that during sliding, the asperity will remain in contact with the substrate for at least 
a distance of the order of a few times the diameter of the contact region. Hence, the frictional heat source will be stationary
(not moving) with respect to the block but sliding on the substrate. Let $f_1(v)$ be the average temperature in the
sliding contact assuming that the entire frictional energy goes into the substrate. Similarly, let $f_2(v)$ be the average temperature
in the stationary contact assuming that the entire frictional energy goes into the substrate. In reality, a fraction $s$ of the total energy
will go into the sliding contact and a fraction $1-s$ into the stationary contact, and the (average) temperature in the two contact regions will
be $s f_1(v)$ and $(1-s)f_2(v)$. Assuming that the (average) temperature is continuous gives
$$s f_1(v) =(1-s) f_2(v)$$ 
or $s= f_2(v)/[f_1(v)+f_2(v)]$. Using this the average flash temperature in the contact will be
$$T_{\rm flash} = s f_1(v) = {f_1(v) f_2(v) \over f_1(v)+f_2(v)}.$$

\begin{figure}
\includegraphics[width=0.47\textwidth,angle=0.0]{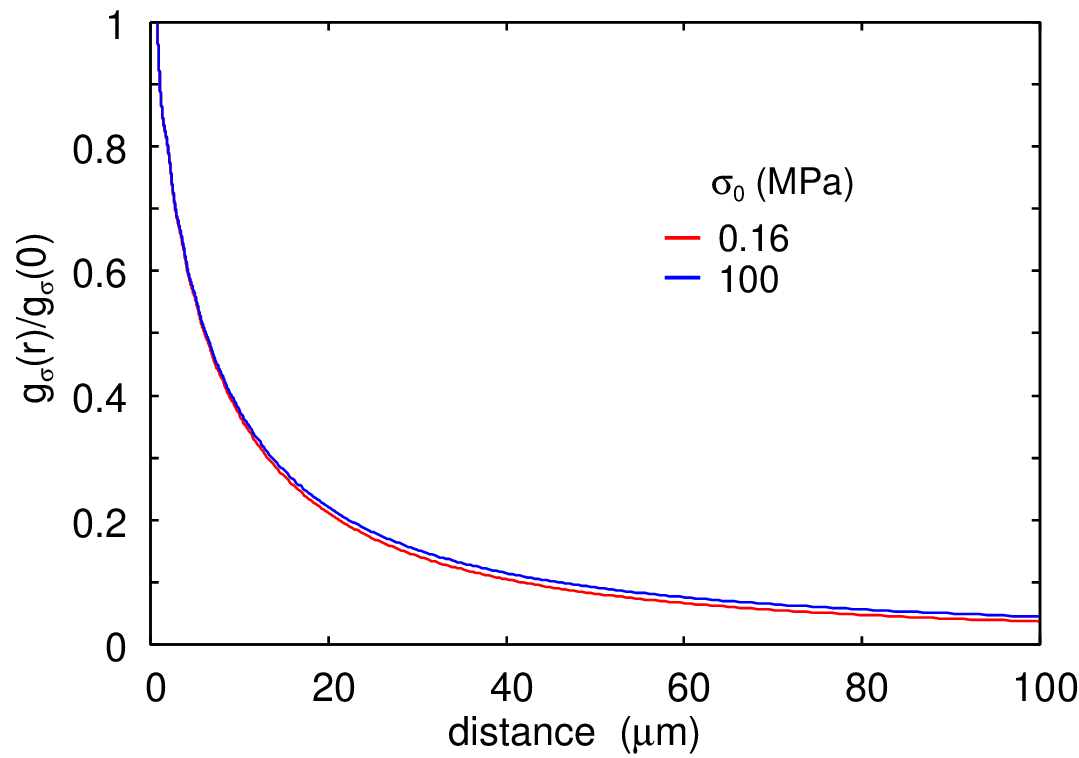}
\caption{\label{1distanceMicrometer.stresscorrelationNormalized.granite.eps}
The normalized stress correlation $g_{\sigma}(r)/g_{\sigma}(0)$ as a function of the separation $r=|{\bf x}-{\bf x}'|$
for the contact between two granite surfaces, using the elastic properties of quartz (which are nearly the same as those for silica). 
The red and blue lines were obtained assuming nominal contact pressures of $0.16$ and $100$ MPa, respectively. 
If the contact diameter $r_0$ were defined by $g_{\sigma}(r_0) = \alpha g_{\sigma}(0)$ 
with $\alpha = 0.2$, then we would get $r_0 \approx 20 \ {\rm \mu m}$. However, the macroasperity contacts 
consist of a central compact region surrounded by 
disconnected islands (see Ref. \cite{Persson3}); these islands of contact 
are elastically correlated with the central part of the stress correlation function 
$g_{\sigma}(r)$, which results in a long tail of $g_{\sigma}(r)$ extending to larger distances $r$.
}
\end{figure}

Fig. \ref{1distanceMicrometer.stresscorrelationNormalized.granite.eps}
shows the normalized stress correlation $g_{\sigma} (r)/g_{\sigma} (0)$ as a function of the separation $r=|{\bf x}-{\bf x}'|$
for the contact between two granite surfaces using the elastic properties of quartz (which are nearly the same as those for  silica).
If the contact radius were defined by $g_{\sigma} (r) = \alpha g_{\sigma} (0)$
with $\alpha = 0.5$, then we would get $r_0 \approx 6 \ {\rm \mu m}$. However, the macroasperity contacts
consist of a central compact region surrounded by 
disconnected islands (see Ref. \cite{Persson3}), and these islands of contact
are elastically correlated with the central part of the stress correlation function
$g_{\sigma} (r)$, which results in a long tail of $g_{\sigma} (r)$ extending to larger distances $r$.

The red and blue lines in
Fig. \ref{1distanceMicrometer.stresscorrelationNormalized.granite.eps} 
were obtained assuming nominal contact pressures of $0.16$ and $100 \ {\rm MPa}$, respectively.
Note that $g_{\sigma} (r)/g_{\sigma} (0)$ is nearly independent of the applied pressure, which holds true as long as the
relative contact area satisfies $A/A_0 \ll 1$. The reason is that increasing the nominal pressure $\sigma_0$ when $A/A_0 \ll 1$ increases
the number of macroasperity contact regions in proportion to $\sigma_0$, while the pressure distribution within the macroasperity contact
regions remains nearly unchanged. As long as $A/A_0 \ll 1$, the elastic coupling between the macroasperity contact regions is small,
and $g_{\sigma} (r)/g_{\sigma} (0)$ will nearly vanish for $r=|{\bf x}-{\bf x}'|$ 
larger than the diameter of the macroasperity contact regions. These statements require a large enough system 
with a surface roughness power
spectrum that has a roll-off, which is always the case for surfaces of engineering interest.

\begin{figure}
\includegraphics[width=0.47\textwidth,angle=0.0]{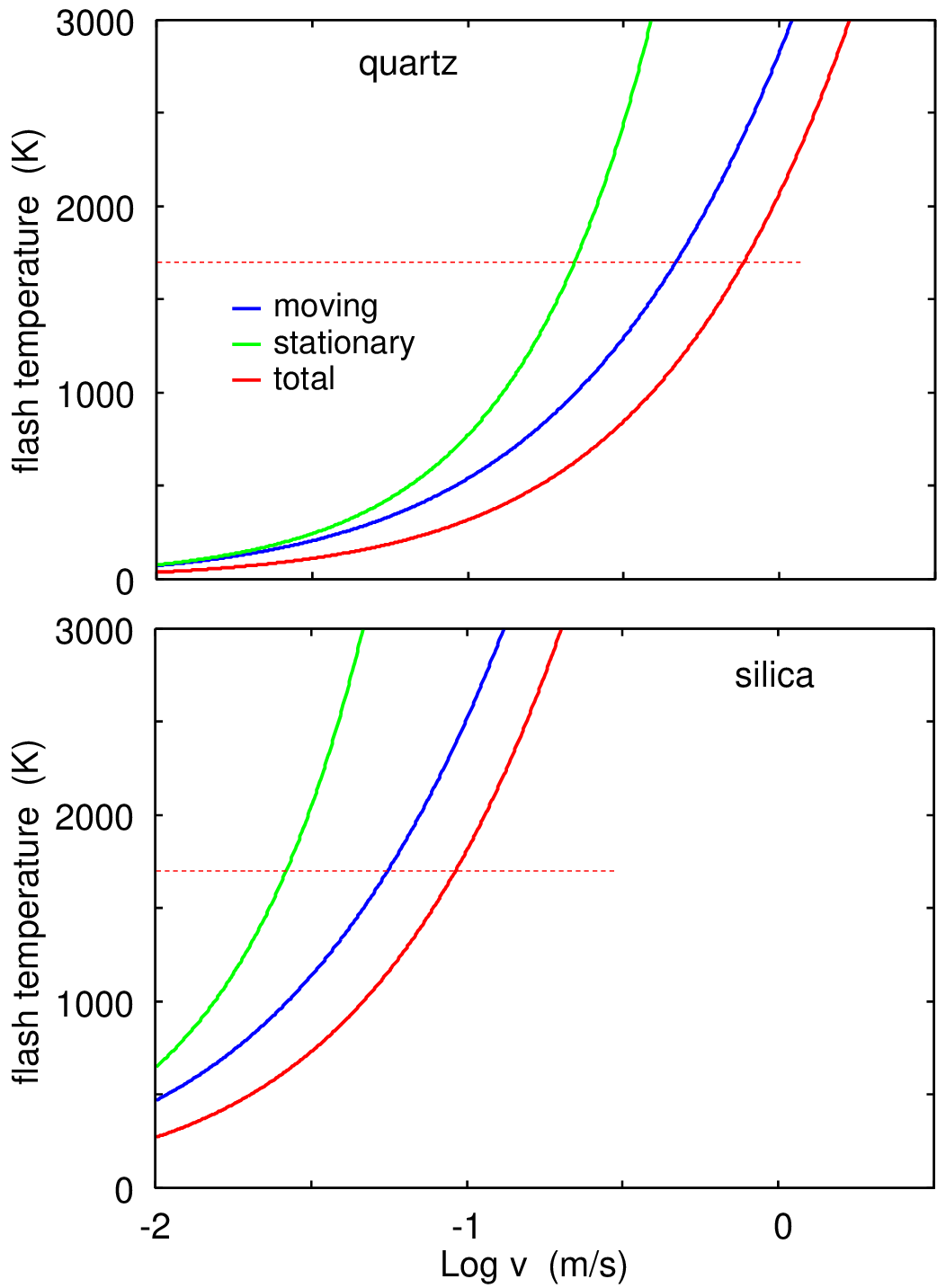}
\caption{\label{1logv.2Temp.quartz.eps}
The temperatures $f_1(v)$ and $f_2(v)$ in the sliding (blue line) and stationary (green line) contacts,
assuming the entire frictional energy goes into the sliding and stationary contact, respectively.
Thus, the red line is the result when part of the frictional energy goes into the sliding contact and part into the
stationary contact in such a way that the temperature is continuous at the contacting interface. The red line is given by
$f_1 f_2/(f_1+f_2)$. The melting temperature of quartz (about $1700 \ {\rm K}$, which is also
the temperature where quartz and silica become identical high-viscosity fluids) is indicated by the dashed line.
For the silica system, the quartz melting temperature is reached for
a sliding $v \approx 0.1 \ {\rm m/s}$.}
\end{figure}

\begin{figure}
\includegraphics[width=0.47\textwidth,angle=0.0]{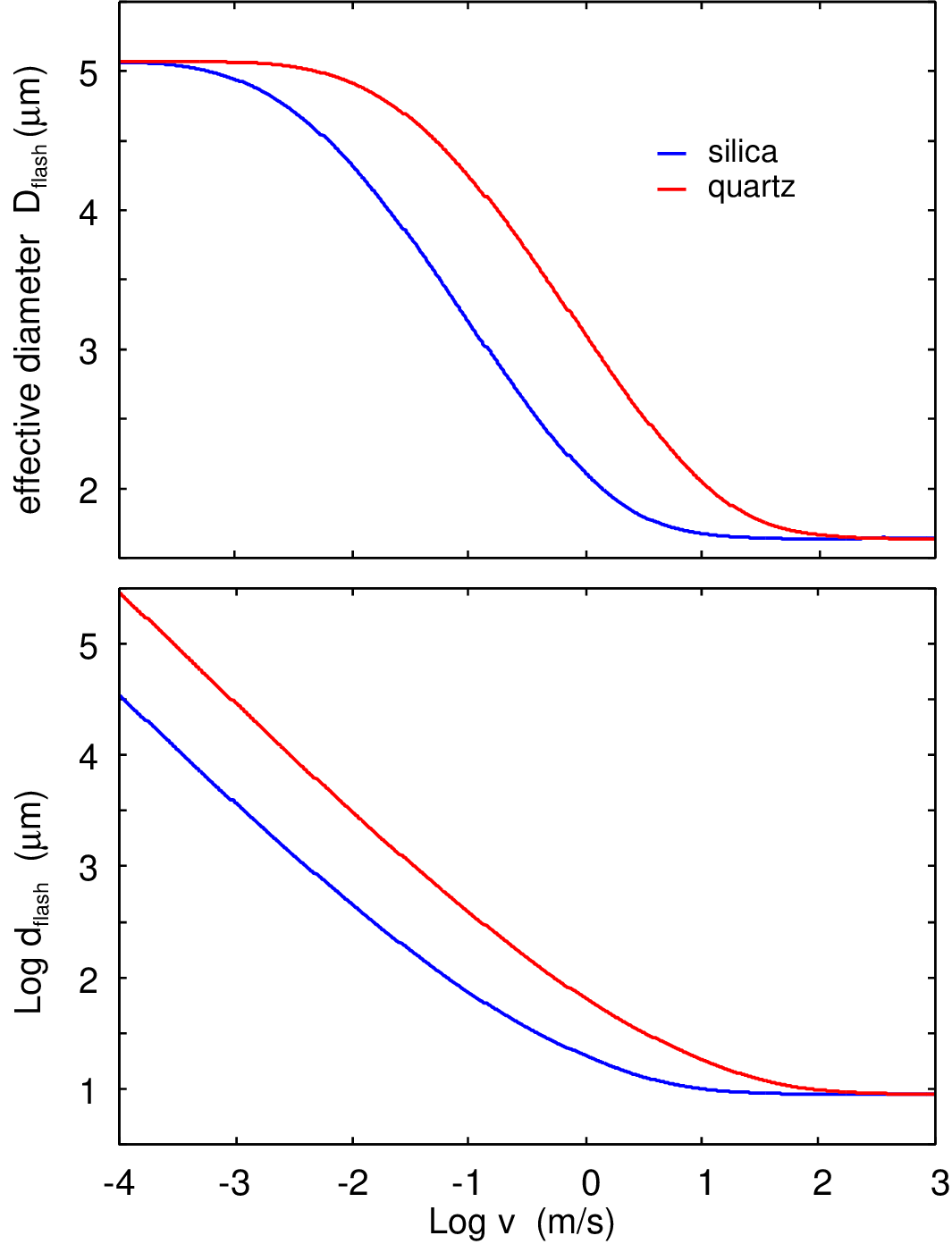}
\caption{\label{1logv.2logToverTprime.silica.quartz.eps}
(a) The effective diameter of the hotspots,  $D_{\rm flash} = [-T_{\rm flash} /\nabla^2 T_{\rm flash} ]^{1/2}$ and
(b) the logarithm of the temperature slope length, $d_{\rm flash} = -T_{\rm flash}/T'_{\rm flash} $ as a function of the sliding speed
(log-log scale). When the slope length $d_{\rm flash}$ is of the order of the effective radius of the macro-asperity contact region 
(about $10 \ {\rm \mu m}$ for the present system)
the temperature at the exit of the temperature profile will be roughly twice the average temperature $T_{\rm flash}$.
%
% The following TEXT was an EXACT COPY of that in the MAIN text. SO I COMMENTED IT OUT.
%
% In the present case for silica, this occurs at a sliding speed of $v \approx 10 \ {\rm m/s}$. However, at this sliding speed 
% silica (and quartz) have melted. At the melting temperature of quartz the decay length is approximately $100 \ {\rm \mu m}$ for both 
% silica and quartz. Hence, the temperature profile exhibits only a relatively small upward ``tilt'' towards the exit side of the 
% contact region, even immediately before melting occurs.
}
\end{figure}

Fig. \ref{1logv.2Temp.quartz.eps} shows the temperatures $f_1(v)$ and $f_2(v)$ in the sliding (blue line) and stationary (green line) contacts,
assuming the entire frictional energy goes into the sliding or stationary contact, respectively.
The red line is the result when part of the frictional energy goes into the sliding contact and part into the
stationary contact in such a way that the (average) temperature is continuous at the contacting interface. The red line is given by
$f_1 f_2/[f_1+f_2]$. For the silica-silica contact, the melting temperature of quartz (about $1700 \ {\rm K}$, which is also
the temperature where quartz and silica become identical high-viscosity fluids) is reached for
a sliding velocity of $\approx 0.1 \ {\rm m/s}$. 
We note that at the high normal and tangential stresses (of order $10 \ {\rm GPa}$) 
acting in the contact regions between the quartz grains on the opposite surfaces 
for granite sliding on granite, the quartz is likely to locally
convert to a silica-like material, which can have a large effect on the local temperature as the thermal
conductivity of silica is a factor of $\sim 1/10$ smaller than that of quartz. This is due to the much shorter phonon
mean free path in the disordered silica compared to quartz.

The melting temperature of quartz is $\approx 1700~^\circ {\rm C}$, while the melting temperature
of silica is not well defined, as it softens gradually with increasing temperature.
If the friction coefficient were velocity-independent and equal to $1$, as assumed above
and found in experiments at room temperature, the temperature in the asperity contact regions
for silica would reach $\approx 1700~^\circ {\rm C}$ at a sliding speed  of 
$\approx 0.1 \ {\rm m/s} \ {\rm m/s}$. However, in reality, the friction coefficient, and hence the shear stress
$\sigma_{\rm f}$, drops strongly at high temperature so that silica would actually melt
at much higher sliding speeds than is predicted for a velocity-independent friction coefficient.

We now focus on $T'_{\rm flash}$ and $\nabla^2 T_{\rm flash}$.
We define the width $D_{\rm flash}$ and the slope-length $d_{\rm flash}$ by
$$D_{\rm flash} = \left [ {-T_{\rm flash} \over \nabla^2 T_{\rm flash}} \right ]^{1/2}\eqno(25)$$
$$d_{\rm flash} = {-T_{\rm flash} \over T'_{\rm flash}}\eqno(26)$$
Both $D_{\rm flash}(v)$ and $d_{\rm flash}(v)$ depend on the sliding speed $v$. 
The interpretation of these quantities are best illustrated in the classical limit of
a single sliding Hertz contact (radius $R$) with the heat source (3). 
We can consider this as a macroasperity contact but where the roughness at shorter length scale
is neglected so resulting from a smooth macroasperity. 
For this one-length scale problem, in Ref. \cite{CompareSteel} we have show that 
$$D_{\rm flash} = \alpha R\eqno(27)$$
where $\alpha \approx 0.63$ for very low velocity (stationary contacts) and $\alpha \approx 0.45$ for very high velocities. 
Here high and low velocities refer to $v \gg v^*$ and $v \ll v^*$, with $v^* = D/R$, respectively.
Similarly, the slope parameter for high sliding speeds
$$d_{\rm flash} \approx 1.36 R \eqno(28)$$
while $d_{\rm flash}=\infty$ for $v=0$. Note that for high sliding speed
$d_{\rm flash} /D_{\rm flash} \approx 3.0$.

In Ref. \cite{CompareSteel} we found that these equations are consistent with the measured flash
temperature for steel sliding on steel\cite{SutterContact}. The reason for this is that the steel deform plastically already at relative long length scales
so that effectively roughness occur only on a relative narrow wavelength region, and in particular 
the short wavelength roughness on the original surfaces is irrelevant for the flash temperature. 
As we will now show, this is not the case for the granite studied above, where the
classical approach fail because of the multiscale roughness. 

In Fig. \ref{1logv.2logToverTprime.silica.quartz.eps} we show (a) the width $D_{\rm flash}$ and (b) the slope length $d_{\rm flash}$ as a function of the sliding speed.
When the slope length $d_{\rm flash}$ is on the order of the effective radius of the macro-asperity contact region 
(about $10 \ {\rm \mu m}$ in the present case)
the temperature at the trailing edge % exit of the temperature profile 
will be roughly twice the average temperature $T_{\rm flash}$.
In the present case for silica, this occurs at a sliding speed of $v \approx 10 \ {\rm m/s}$. However, at this sliding speed, the
silica (or quartz) has melted. At the melting temperature of quartz, the decay length is approximately $100 \ {\rm \mu m}$ for both 
silica and quartz. Hence, the temperature profile exhibits only a relatively small upward tilt towards 
the trailing edge, % the exit side of the contact region, 
even immediately before melting occurs.

Fig. \ref{1logv.2logToverTprime.silica.quartz.eps}(a) shows that the width $D_{\rm flash}$ decreases by a factor of
$1.649/5.056 \approx 0.33$ as the sliding speed increases. This differ from the result for a sliding Hertz contact where
$D_{\rm flash}$ decreases by a factor of $0.45/0.63 \approx 0.71$ as the sliding speed increases.

Next let us focus on $d_{\rm flash}$. As in the classical case $1/d_{\rm flash} \rightarrow 0 $ as $v\rightarrow 0$.
For large speeds $d_{\rm flash} \approx 9.00 \ {\rm \mu m}$ and $D_{\rm flash}\approx 1.649  \ {\rm \mu m}$ 
giving $d_{\rm flash}/D_{\rm flash}\approx 5.45$ while the classical description of sliding 
Hertz contact predict $d_{\rm flash}/D_{\rm flash}\approx 3.0$.
It follows that the classical description fails to accurately describe the flash temperature.

\begin{figure}
\includegraphics[width=0.47\textwidth,angle=0.0]{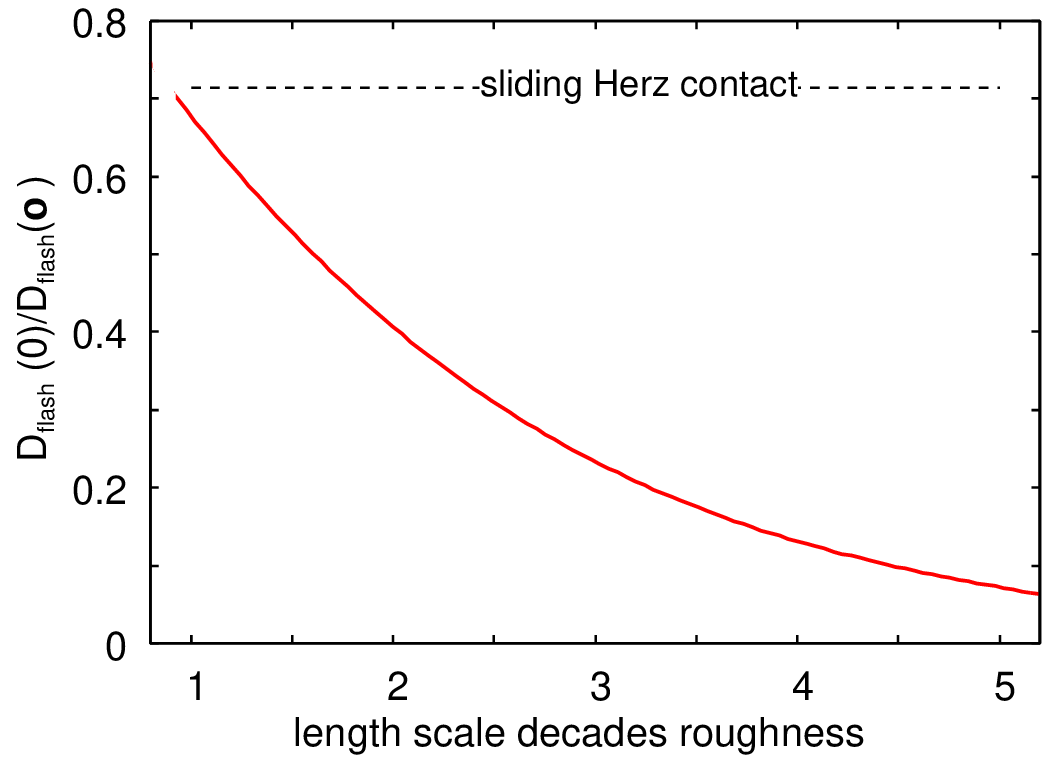}
\caption{\label{1decade.2ratioD.eps}
The ratio $D_{\rm flash}(0)/D_{\rm flash}(\infty)$ between $D_{\rm flash}(v)$ for $v=0$ and $\infty$,
as a function of the number of decades of roughness
if a self-affine fractal surface with the Hurst exponent $H=0.8$.
}
\end{figure}

\begin{figure}
\includegraphics[width=0.47\textwidth,angle=0.0]{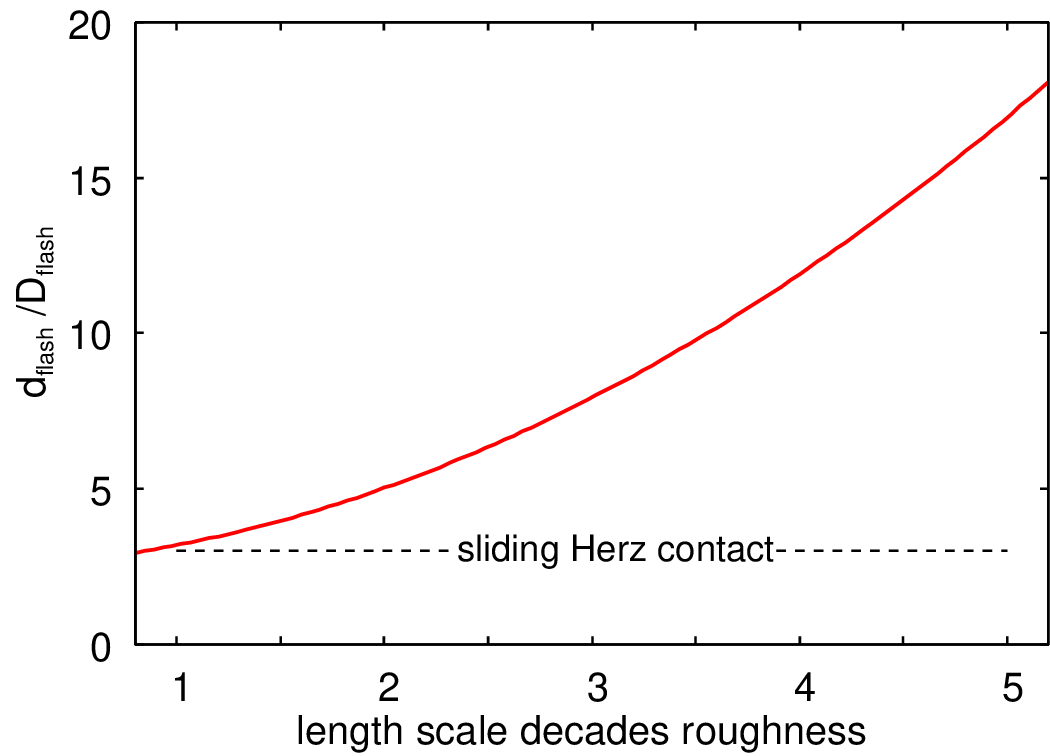}
\caption{\label{1decades.2ratio.eps}
The ratio $d_{\rm flash}(\infty)/D_{\rm flash}(\infty)$ as a function of the number of decades of roughness
of a self-affine fractal surface with the Hurst exponent $H=0.8$.
}
\end{figure}

To further illustrate the failure for the classical theory to describe correctly the flash 
temperature we show in Fig. \ref{1decade.2ratioD.eps} the ratio $D_{\rm flash}(0)/D_{\rm flash}(\infty)$
and in Fig. \ref{1decades.2ratio.eps} the ratio $d_{\rm flash} (\infty)/D_{\rm flash}(\infty)$ as a function of the
number of decades in roughness for a serface with self affine fractal roughness with the Hurst exponent $H=0.8$.
The dashed lines gives the classical result for sliding Hertz contact. The figures shows that the
sliding Hertz theory gives approximately correct results for surfaces with roughness over 1 decade in length
scale but fail severely when the roughness occur over 2 or more decades of roughness.

We will now comment on the assumption that the heat source $\dot q ({\bf x})$  is proportional to the local
stress $\sigma ({\bf x})$. Assume that two solids are squeezed into contact with the nominal stress $\sigma_0$.
If we study the interface at the magnification $\zeta$ we observe the normal stress 
$\sigma ({\bf x})$. At the highest magnification $\zeta_1$, where the atomic-scale roughness can be observed, the stress
$\sigma ({\bf x}) = \sigma_1 ({\bf x})$ is the true normal stress acting at the interface
but this stress is in general not linearly related to the frictional shear stress. However, if we study the 
interface at a lower magnification where the apparent relative contact area $A({\bf x})/A_0 \lesssim 0.3 $
then contact mechanics theories for elastic and elastoplastic contact
between randomly rough surfaces show that  $A ({\bf x})/A_0$ depends linearly
on $\sigma ({\bf x})$. Hence, at this lower magnification
we expect the frictionals shear stress $\sigma_{\rm f} ({\bf x})$ to be proportional to $\sigma ({\bf x})$, 
and the relation $\sigma_{\rm f} ({\bf x})=\mu \sigma ({\bf x})$ to hold. 
In this case $\dot q ({\bf x})$  is proportional to $\sigma ({\bf x})$.

\begin{figure}
\includegraphics[width=0.47\textwidth,angle=0.0]{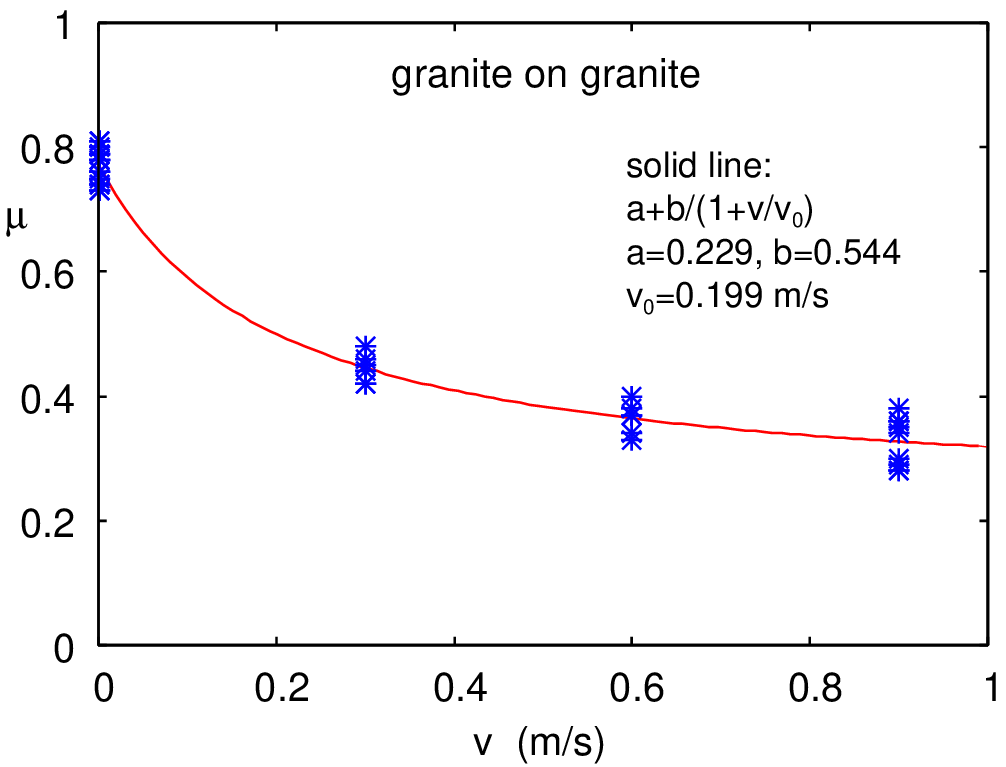}
\caption{\label{1v.2mu.granite.eps}
The plus symbols (+) are measured friction coefficients (from Ref. \cite{earth1}), 
and the solid line represents a fit to the data of the form 
$\mu(v) = a+b/(1+v/v_0)$, with $a=0.229$, $b=0.544$, and $v_0 = 0.199 \ {\rm m/s}$.
}
\end{figure}

\begin{figure}
\includegraphics[width=0.47\textwidth,angle=0.0]{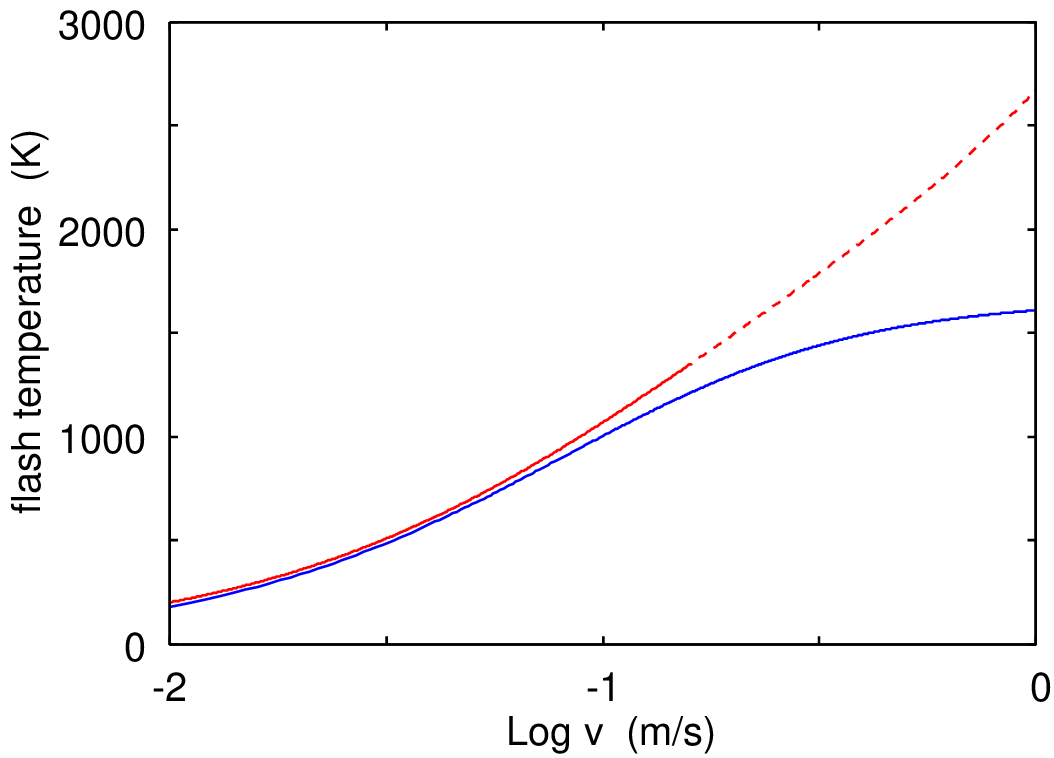}
\caption{\label{MELT.1logv.2Temp.both.eps}
Blue line: the flash temperature calculated from Eqs. (21) and (22). Red line: the theoretically 
calculated flash temperature using the measured friction coefficient.
}
\end{figure}

\vskip 0.2cm
{\bf 6 Application to earthquake dynamics}

Geophysical observations often reveal large dynamic stress drops during earthquakes, indicating
that the effective dynamic friction coefficient between the fault surfaces is
substantially lower than the static or breakloose friction coefficient. During major seismic events, 
the slip velocities at the fault line are typically of the order of $\sim 1 \ {\rm m/s}$. Laboratory
friction studies of granite sliding on granite demonstrate that the friction coefficient drops from $\mu \approx 0.8$
for very low sliding speeds, to $\approx 0.35$ at speeds of the order of $\sim 1 \ {\rm m/s}$. 
It has been proposed that the origin of this drop in the friction coefficient is flash heating
in the contact regions\cite{earth2}. 

Barbery et al.~\cite{earth1} have measured the velocity dependence of the friction between Westerly granite
under a normal pressure of $\approx 9 \ {\rm MPa}$. The sliding blocks were initially ground flat, but no information
about the surface roughness was provided.
The resulting friction coefficient measured for granite sliding on granite is shown
in Fig. \ref{1v.2mu.granite.eps}. The solid fit-line is given by
$$\mu(v) = a+{b\over 1+v/v_0}, \eqno(29)$$
where $a=0.229$, $b=0.544$ and $v_0 = 0.199 \ {\rm m/s}$.  We assume that in the studied velocity interval
the velocity dependency of the friction coefficient is due to the increase in the flash temperature. 
Here, we use a similar temperature dependency of the friction coefficients as found from studies of friction on ice:
$$\mu = \mu_0 \left (1-{T\over T_{\rm m}} \right )^\alpha, \eqno(30)$$
where $T_{\rm m}$ is the melting temperature of quartz, which we take to be $T=1700 \ {\rm K}$ and
where $\mu_0 = a+b = 0.843$ is the static or breakloose friction coefficient. 
%For ice friction the exponent
%$\alpha$ was found to be $\approx 0.15$ for ice on ice\cite{ice1} but approximately $0.6$ for silica sliding on ice\cite{ice2}.
Here, we use $\alpha = 0.3$. By combining (29) and (30), we obtain the velocity dependence of the flash temperature
shown in Fig. \ref{MELT.1logv.2Temp.both.eps} as the blue line. The red line is the calculated flash
temperature using the theory presented above for silica, and is obtained from the red curve in Fig. \ref{1logv.2Temp.quartz.eps}(b)
by scaling it with the $\mu(v)$ given by (29). Here, we have used that the flash temperature is proportional to
$\mu(v)$ and that the curves in Fig. \ref{1logv.2Temp.quartz.eps}(b) were obtained with $\mu(v)=1$.
We note that at the high normal and shear stresses (and high temperatures) occurring in the contact regions between 
quartz grains, quartz is most likely locally converted into silica \cite{rpp}.

The theory for the flash temperature is, of course, not valid when the temperature exceeds the melting temperature.
In fact, even before reaching the melting temperature is reached, the silica softens and the contact radius of the macroasperities
increases, while the average stress $\bar \sigma$ decreases at least as fast as $1/r_0^2$. As a result, the flash temperature becomes smaller
than it would be if this effect were neglected [see (19) where the prefactor $\bar \sigma _0 r_0 \sim 1/r_0$]. This can explain
why the flash temperature deduced from the measured friction coefficient flattens out as the quartz melting temperature is approached.

\vskip 0.2cm
{\bf 7 Summary and conclusion}

We have presented the first analytical theory for the flash temperature which includes the surface roughness
on arbitrary many decades in length scale.
The most important results is (14), which gives the weighted 
average temperature increase in the contact regions. We note that the weighting factor $\sigma ({\bf x})$
is proportional to the heat source so the (14) is the optimum average which can be defined. In addition
the theory gives information about the variation of the temperature within the macroasperity contact regions
as contained in derivatives of the temperature profile [equations (16) and (17)]. 
We have shown that for roughness on a single length scale the theory reduces to the 
classical theory of the flash temperature by Jaeger, Archard and Greenwood for circular moving heat sources.
However, for surfaces with multiscale roughness the classical theories for the flash temperature
fail severely.

\vskip 0.2cm
{\bf Appendix A}

\begin{figure}
\includegraphics[width=0.47\textwidth,angle=0.0]{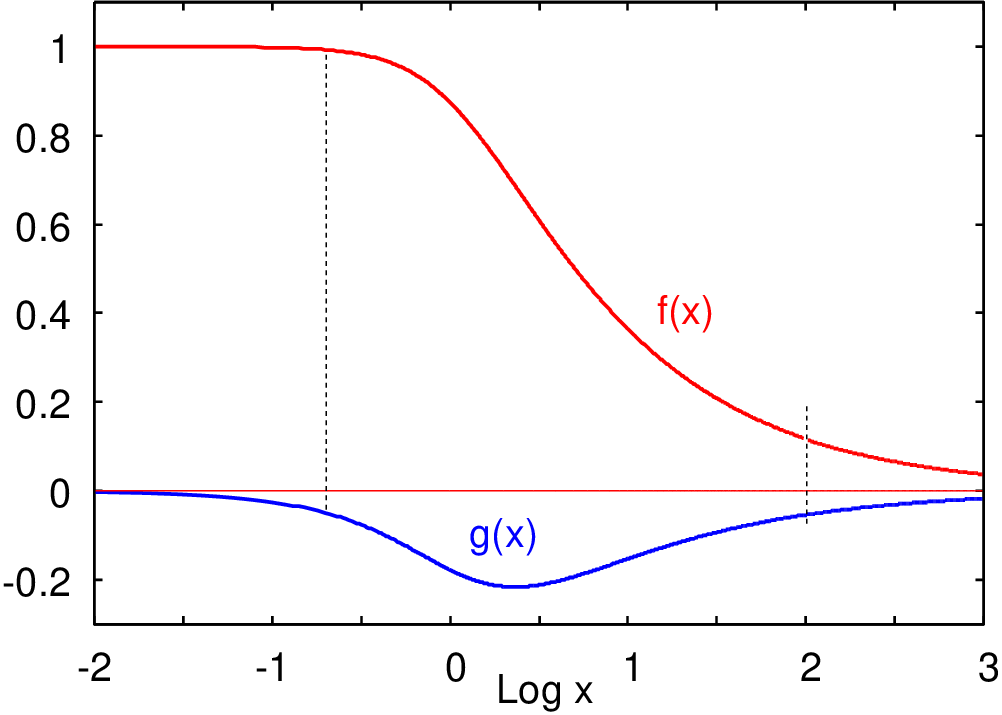}
\caption{\label{1logx.2g.and.f.integrals.eps}
The function $f(x)$ and $g(x)$ given by (A1) and (A4). The region to the left of the first dashed vertical line is given by the
asymptotic small-$x$ results (A2) and (A5), and the region to the right of the second vertical line is given by the asymptotic large-$x$
results (A3) and (A6) where we have used $\epsilon=0.03$. The region in between are given by the integrals (A1) and (A4) by numerical integration. 
}
\end{figure}

\begin{widetext}
Consider the integral
$$f(x) = {2\over \pi} {\rm Re} \int_0^{\pi/2} d\phi \, {1\over (1-ix \, {\rm sin} \phi)^{1/2}}\eqno(A1)$$
As $x \rightarrow 0$ we get 
$$f(x) \approx 1-{3 x^2 \over 16}\eqno(A2)$$
We are also interested in the limit $x \rightarrow \infty$. Assume $\epsilon \ll 1$. For $\phi < \epsilon$ we have
${\rm sin}\phi \approx \phi$ and we can write
$$f(x) \approx {2\over \pi} {\rm Re} \left [ \int_0^\epsilon d\phi \, {1\over (1-ix \, \phi)^{1/2}}+ \int_\epsilon^{\pi/2} d\phi \, {1\over (1-ix \, {\rm sin} \phi)^{1/2}}\right ]$$
For $x \gg 1/\epsilon$ in the second term in this expression we can replace $(1-ix \, {\rm sin} \phi)^{1/2}$ with $(-ix \, {\rm sin} \phi)^{1/2}$ to get
$$f(x) \approx {2\over \pi} {\rm Re} \left [ {2\over ix} \left (1-(1-ix \, \epsilon)^{1/2}\right ) 
+ {1\over (-ix)^{1/2}} \int_\epsilon^{\pi/2} d\phi \, {1\over ({\rm sin}\phi )^{1/2}} \right ]\eqno(A3)$$

Next, consider the integral
$$g(x) = {2\over \pi} {\rm Re} \int_0^{\pi/2} d\phi \, {i \, {\rm sin}\phi \over (1-ix \, {\rm sin} \phi)^{1/2}}\eqno(A4)$$
As $x \rightarrow 0$ we get
$$g(x) \approx -{x\over 4}\eqno(A5)$$
We are also interested in the limit $x \rightarrow \infty$. For $x \gg 1/\epsilon$ we get
$$g(x) \approx {2\over \pi} {\rm Re} \left [ \int_0^\epsilon d\phi \, {i \, \phi \over (1-ix \, \phi)^{1/2}}+ {1\over (-ix)^{1/2}} \int_\epsilon^{\pi/2} d\phi \, {i \, {\rm sin}\phi \over ({\rm sin} \phi)^{1/2}}\right ]$$
$$={2\over \pi} {\rm Re} \left [ {2\over i x^2} \left ({2\over 3}+{1\over 3} (1-ix \, \epsilon)^{3/2}- 
(1-ix \, \epsilon)^{1/2} \right )+{i\over (-ix)^{1/2}} \int_\epsilon^{\pi/2} d\phi \,({\rm sin} \phi)^{1/2}\right ]\eqno(A6)$$

Fig. \ref{1logx.2g.and.f.integrals.eps} shows the integrals (A1) and (A4).
The region to the left of the first dashed vertical line is given by the
asymptotic small-$x$ results (A2) and (A5), and the region to the right of the second vertical line is given by the asymptotic large-$x$
results (A3) and (A6) where we have used $\epsilon=0.03$. The region in between are given by the integrals (A1) and (A4) by numerical integration. 

\end{widetext}

\end{document}